\pdfoutput=1
\documentclass[a4paper,reprint,twocolumn,notitlepage,aip,nofootinbib]{revtex4-2}
\usepackage[left=1.5cm,right=1.5cm,top=1cm,bottom=1.5cm,includeheadfoot]{geometry}
\usepackage[english]{babel}
\usepackage[T1]{fontenc}
\usepackage{tgtermes} 
\usepackage{tgheros} 

\usepackage{amssymb}
\usepackage{amsmath}
\usepackage{amsthm}
\usepackage{bm} 
\usepackage{bbm}

\usepackage{graphicx}
\usepackage[font=small]{caption}
\usepackage{subcaption}
\usepackage{ushort}
\PassOptionsToPackage{hyphens}{url} 
\usepackage[breaklinks=true]{hyperref}
\bibpunct{[}{]}{;}{n}{}{} 
\usepackage[dvipsnames]{xcolor}
\usepackage{tikz}
\usetikzlibrary{3d}

\usepackage{float}
\makeatletter
\let\newfloat\newfloat@ltx
\makeatother
\usepackage{algorithm}
\usepackage{algpseudocode}

\theoremstyle{definition}

\renewcommand{\H}{\mathcal{H}}
\newcommand{\M}{\mathcal{M}}
\newcommand{\crit}{\mathcal{C}}
\newcommand{\Bcrit}{\mathcal{B}_\mathrm{c}}
\newcommand{\R}{\mathbb{R}}
\newcommand{\C}{\mathbb{C}}
\newcommand{\e}{\mathrm{e}}
\renewcommand{\i}{\mathrm{i}}
\renewcommand{\d}{\,\mathrm{d}} 
\newcommand{\id}{{\hat I}}
\newcommand{\bb}{{\bm{b}}}
\newcommand{\bbeta}{{\bm{\beta}}}
\newcommand{\rrho}{{\bm{\rho}}}
\newcommand{\Gram}{\mathcal{G}}
\newcommand{\uu}{\bm{u}}
\newcommand{\vv}{\bm{v}}
\newcommand{\ww}{\bm{w}}
\newcommand{\Bdom}{\mathcal{B}}
\newcommand{\gmap}{{\bm{g}}}

\DeclareMathOperator{\interior}{int}
\DeclareMathOperator{\im}{im}
\DeclareMathOperator{\Var}{Var}
\renewcommand{\Re}{\mathop\mathsf{Re}}

\newcommand{\eps}{\varepsilon}

\begin{document}

\title{Constrained Search in Imaginary Time}

\author{Markus Penz}
\email{m.penz@inter.at}
\affiliation{Max Planck Institute for the Structure and Dynamics of Matter and Center for Free-Electron Laser Science, Hamburg, Germany}
\affiliation{Department of Computer Science, Oslo Metropolitan University, Oslo, Norway}

\author{Robert van Leeuwen}
\affiliation{Department of Physics, Nanoscience Center, University of Jyv\"askyl\"a, Finland}

\begin{abstract}
We introduce an imaginary-time evolution method to evaluate the pure-state constrained-search functional from density-functional theory formulated on finite lattices. Simultaneously, it yields a potential that produces a prescribed density of an eigenstate. Besides being a computational scheme, this allows one to gain theoretical insights into the density-potential mapping. The method can be generalized to the optimization of the expectation value of a general self-adjoint operator on a finite-dimensional Hilbert space under a finite number of expectation-value constraints for commuting self-adjoint operators.
\end{abstract}
\maketitle


\section{Introduction}

Variational principles play a principal role in many areas of physics and are often realized as a constrained-optimization problem. One prime example is the ground-state problem of quantum mechanics, where the expectation value of the Hamiltonian is minimized over all Hilbert-space vectors $\Psi\in\H$ under the constraint $\|\Psi\|=1$. Density-functional theory (DFT)~\cite{eschrig2003-book,vonBarth2004basic,burke2007abc,dreizler2012-book}, a standard approximation method for the many-body ground-state problem widely employed in chemistry and materials science, is built around the concept of the `constrained-search functional'~\cite{percus1978,Levy79,Lammert2007,Penz2023-Review-Part-I}, which is defined as the optimization of the expectation value of the internal part of the Hamiltonian under a fixed-density constraint.
This delegates the complexity of the ground-state problem to this density functional, for which practical approximations are available in quantum chemistry~\cite{toulouse2022review}. Nevertheless, for calculating the functional values themselves, the complexity of the full Hilbert space remains and this makes it notoriously hard to evaluate the value of the functional at a given density.
For this reason, any new optimization method that can be applied to this problem is of great interest. Such a method that achieves its goal by imaginary-time evolution is proposed and studied in this work.

Imaginary-time evolution itself is a well-established method that uses Wick rotation $t = -\i\tau$ of the time-dependent Schrödinger equation to arrive at a diffusion equation
\begin{equation}
    -\partial_\tau\Psi = \hat H\Psi
\end{equation}
that describes evolution toward the ground state of the time-independent Hamiltonian $\hat H$ as $\tau\to\infty$. For an initial state $\Psi_0$ expanded in an eigenbasis $\{\Phi_k\}$ with eigenvalues $\{\eps_k\}$ of $\hat H$ this evolution means that
\begin{equation}
    \Psi(\tau) = \e^{-\tau\hat H}\Psi_0 = \e^{-\tau\hat H} \sum_k c_k\Phi_k = \sum_k c_k \e^{-\tau\eps_k} \Phi_k,
\end{equation}
and consequently, eigenstates with smaller energies get amplified compared to those with larger energies. Note that this evolution is not unitary, so the normalization of $\Psi$ is not conserved. If the wavefunction is normalized after each time step, then over (imaginary) time only the eigenstate with the smallest energy and $c_k\neq 0$ survives. The same effect of normalization can be established by shifting the Hamiltonian by $-\eps_0\hat I$, which has the effect of a Lagrangian multiplier for the normalization constraint.
While this method has been known as a numerical tool for solving ground-state problems for a long time~\cite{Goldberg1967-I,Goldberg1967-II,Kosloff1986} and was also applied to the path integral technique~\cite{Grimm1969,Lawande1969,Skinner1985}, the inclusion of additional constraints directly into the method, as far as we know, was previously not taken into account.

The novelty of our method consists in the possible consideration of multiple constraints of type $\langle\hat B_i\rangle_\Psi = b_i \in \R$, where the $\hat B_i$ can be arbitrary self-adjoint operators that are linearly independent and that mutually commute. Yet, for the sake of clarity, the method will first be described with density constraints in the context of DFT, as an interesting and useful application, while the general formulation together with a whole extension of DFT to generalized settings will be laid out in Appendix~\ref{app:gft}.
While there are other constrained-optimization methods, like the Blahut--Arimoto algorithm applied to density matrices~\cite{hayashi2024-Arimoto-Blahut} or versions of Riemannian optimization~\cite{smith2014optimization} that could be formulated on a submanifold of states that fulfill the given constraints, we do not know of any comparable method that relies exclusively on imaginary-time evolution. The method thus has a clear conceptual basis and a nice interpretation in terms of a evolution path on a constraint manifold in state space.
As a byproduct, the method yields the necessary potential in order to produce exactly the constraints in a ground state of the Schrödinger equation, thus also solving the inverse Kohn--Sham problem of DFT~\cite{Shi2021}.
We highlight that the method is not only of interest for the field of DFT and is not even limited to quantum mechanics, since it can be formulated even more generally as minimzing a cost function $\langle\hat A\rangle_\Psi$ under the constraints $\langle\hat B_i\rangle_\Psi = b_i$. 

In order to have all steps of the procedure well-defined, we restrict ourselves to finite-dimensional Hilbert spaces. Such spaces conventionally arise in quantum chemistry through the use of finite basis sets and in the study of finite model systems, such as the Hubbard model~\cite{carrascal2015hubbard,Qin2022}, or in finite-differencing approaches to the Schr\"odinger equation~\cite{Kohn1983,CCR1985,Farzanehpour2012,Nielsen2018}. The optimization method itself will first be formulated for spinless particles on a finite lattice in Section~\ref{sec:opt-method}, while a generalization to arbitrary other models is given in Appendix~\ref{app:gft}.
Due to the possibility of getting stuck at excited states, if no further measures are taken, it has a clear connection also to excited-state DFT~\cite{Goerling1999,Giarrusso2023,Yang-Ayers2024}. To still reach an optimum, we overcome the slowdown of the progression of the method close to excited states by adding discontinuous jumps to the otherwise continuous procedure. Section~\ref{sec:algo} explains how the algorithm can be implemented and gives several examples of many-particle lattice systems. We summarize and give an outlook on possible further developments and open problems in Section~\ref{sec:sum}. Finally, four appendices embed the method into a more general context and give mathematical details for results that are used in the formulation of the method.

\section{Optimization method}
\label{sec:opt-method}

\textbf{Quantum particles on a lattice:} We will formulate the method for the setting of finite-lattice DFT~\cite{penz2021-Graph-DFT}.
On a lattice with $M$ sites, define the density operators $\hat\rho_i$ for each lattice-site index $i=1,\ldots,M$. Note that they fulfill $\sum_{i=1}^M\hat\rho_i=N\id$, with $N$ the number of particles and $\hat I$ the identity operator. Consequently, by leaving out one density operator, say the last, the set of operators $\{\hat I, \hat\rho_1, \ldots, \hat\rho_m\}$, $m:=M-1$, becomes linearly independent. The Hamiltonian for a given potential $\vv\in\R^m$ that acts on each lattice site except the last is
\begin{equation}
    \hat H(\vv) := \hat H_0 + \sum_{i=1}^m v_i \hat \rho_i,
\end{equation}
with a fixed internal part $\hat H_0$.
For any normalized state $\Psi\in\H$ in an $L$-dimensional Hilbert space ($L = {M \choose N}$ for spinless fermionic particles; the inclusion of spin is no real obstacle~\cite{Penz2024-Perspective}), the $\rho_i=\langle\hat \rho_i\rangle_\Psi$ are then the occupation numbers on the lattice, where on the last site it holds that
\begin{equation}
    \rho_{M} = N - \sum_{i=1}^m \rho_i
\end{equation}
from the normalization of the density to $N$ particles. Since this fixes $\rho_{M}$, it was not necessary to include it into the description in the first place, a fact that shows up on the dual side of the potential $\vv\in\R^m$ in the form of a gauge fixing $v_M=0$ on the last site. The set of all possible densities (all, as we will see in Appendix~\ref{app:gft}, representable by pure states) is thus
\begin{equation}
    \Bdom = \left\{\bm{\rho}\in\R^m \;\middle|\; 0\leq \rho_i\leq 1, N-1\leq \sum_{i=1}^m \rho_i \leq N \right\}.
\end{equation}
For $M=4$ and $N=2$, this cuts out an octahedron from a cube (Figure~\ref{fig:octahedron}), and in the general case it gives an $(M,N)$-hypersimplex~\cite{penz2021-Graph-DFT}. We show in Appendix~\ref{app:gft} that $\Bdom$ is also the convex hull of the densities from the elements of the orthonormal basis $\{\Phi_k\}$ of $\H$ that diagonalizes all $\hat\rho_i$ simultaneously. This means the $L={M \choose N}$ different densities coming from $\Phi_k$ are exactly the vertices of the convex polytope $\Bdom\subseteq\R^m$.

\begin{figure}[ht]
  \centering
  \resizebox{\columnwidth}{!}{%
  \begin{tikzpicture}
    \coordinate (C1) at (3.37,-0.13);
    \coordinate (C2) at (0.07,-3.12);
    \coordinate (C3) at (4.24,-2.32);
    \coordinate (C4) at (6.67,-3.65);
    \coordinate (C5) at (2.50,-4.42);
    \coordinate (C6) at (3.36,-6.61);
    \foreach \m in {1,...,6}
    	\node[circle,fill,inner sep=1.0pt] (N\m) at (C\m) {};
    \draw (N1) -- (N2);
    \draw[dashed] (N1) -- (N3);
    \draw (N1) -- (N4);
    \draw (N1) -- (N5);
    \draw[dashed] (N2) -- (N3);
    \draw[dashed] (N3) -- (N4);
    \draw (N2) -- (N5);
    \draw (N4) -- (N5);
    \draw (N2) -- (N6);
    \draw[dashed] (N3) -- (N6);
    \draw (N5) -- (N6);
    \draw (N4) -- (N6);
    \node [above] at (N1) {$(1,0,1,0)$};
    \node [left] at (N2) {$(0,1,1,0)$};
    \node [right] at (N4) {$(1,0,0,1)$};
    \node [right] at (N3) {$(0,0,1,1)$};
    \node [left] at (N5) {$(1,1,0,0)$};
    \node [below] at (N6) {$(0,1,0,1)$};
  \end{tikzpicture}
  }
\caption{The density domain $\Bdom$ for $M=4$ and $N=2$ forms an octahedron, with the vertex densities having the two particles localized at two lattice sites~\cite{penz2021-Graph-DFT}.}
\label{fig:octahedron}
\end{figure}

\textbf{Constrained optimization problem:} 
On this finite-dimensional complex Hilbert space $\H$, let $\hat H_0$ be the self-adjoint operator that represents the internal part of the Hamiltonian and stands for the \emph{objective} of the optimization problem:
\begin{equation}\label{eq:opt-problem}
\begin{aligned}
    &\min_{\Psi\in\H} \langle\hat H_0\rangle_\Psi  \quad\text{under the constraints}\\
    &\langle \hat I \rangle_\Psi = \|\Psi\|^2 = 1 \quad\text{and}\\
    &\langle \hat\rho_i \rangle_\Psi = \rho_i \;\;\text{for all}\;\; i=1,\ldots,m.
\end{aligned}
\end{equation}
In shorthand notation, employing the constraint set
\begin{equation}\begin{aligned}
    \M_\rrho := \{ &\Psi\in\H \mid \|\Psi\|=1, \\
    &(\langle\hat\rho_1\rangle_\Psi,\ldots,\langle\hat\rho_m\rangle_\Psi) = (\rho_1,\ldots,\rho_m) = \rrho \} 
\end{aligned}\end{equation}
for given $\rrho\in\Bdom$ this is equivalent to
\begin{equation}
    \tilde F(\rrho) := \min_{\Psi\in\M_\rrho} \langle\hat H_0\rangle_\Psi,
\end{equation}
which exactly defines the pure-state constrained-search functional from DFT. If one is further interested in the convex universal functional $F(\rrho)$, this can be subsequently obtained by forming the convex hull of $\tilde F(\rrho)$~\cite[Prop.~18]{penz2021-Graph-DFT}. In the process of optimization, the method also yields a representing $\vv\in\R^m$ as a byproduct, although the representation will not necessarily be in terms of a ground state of $\hat H(\vv)$ but it could be an excited state instead.

\textbf{Description of the method:} Perform an imaginary-time evolution with the generator
\begin{equation}\label{eq:generator}
    \hat G(\tau) := \hat H(\tau) + v_0(\tau)\id,\quad \hat H(\tau) := \hat H_0 + \sum_{i=1}^m v_i(\tau) \hat\rho_i
\end{equation}
and any initial state $\Psi_0\in\H$ that fulfills the constraints of Eq.~\eqref{eq:opt-problem} (can be found by solving Eq.~\eqref{eq:constraints-in-basis-2}, see below), and where at each time the $v_i(\tau)\in\R$, $i=0,\ldots,m$, are chosen such that the constraints are further secured. This means solving the non-autonomous evolution equation
\begin{equation}\label{eq:evolution}
    -\partial_\tau\Psi(\tau)=\hat G(\tau)\Psi(\tau). 
\end{equation}
The fulfillment of the constraints then demands for all $i=0,\ldots,m$ (denoting $\tau$-dependence will be suppressed from here on)
\begin{equation}\label{eq:constraints-beta}
\begin{aligned}
    0 &= \partial_\tau \langle \hat\rho_i\rangle_\Psi = \partial_\tau \langle\Psi, \hat\rho_i\Psi\rangle \\
    &= -\langle\Psi, \hat G \hat\rho_i \Psi \rangle - \langle\Psi, \hat\rho_i \hat G \Psi \rangle
    =  -\langle\{\hat G, \hat\rho_i\}\rangle_\Psi \\
    &= -\langle\{\hat H_0, \hat\rho_i\}\rangle_\Psi - \sum_{j=0}^m  v_j \langle \{ \hat\rho_j, \hat\rho_i \} \rangle_\Psi \\
    &= -\langle\{\hat H_0, \hat\rho_i\}\rangle_\Psi - \sum_{j=0}^m  v_j (\langle \hat\rho_i\Psi, \hat\rho_j\Psi \rangle + \langle \hat\rho_j\Psi, \hat\rho_i\Psi \rangle).
\end{aligned}
\end{equation}
Here, we used that the $\hat G$ and $\hat\rho_i$ are self-adjoint, set $\hat\rho_0=\hat I$, and introduced the anti-commutator $\{\cdot,\cdot\}$. At this point, it must be stressed that this anti-commutator is the consequence of \emph{imaginary}-time evolution, whereas in \emph{real} time it would be the usual commutator of the Heisenberg equation. For this reason, existing proofs from time-dependent DFT cannot be simply ported to solutions from imaginary-time evolution~\cite{flamant2019}.
It is also only with imaginary time that an equation for $\vv$ can be found like this in first order, while in real time a second-order time derivative is necessary~\cite{Ruggenthaler2015-TDDFT}.
Using further that all $\hat\rho_i$ commute, Eq.~\eqref{eq:constraints-beta} can be simplified to
\begin{equation}\label{eq:constraints-beta-simple}
     \sum_{j=0}^m v_j \langle \hat\rho_i\Psi, \hat\rho_j\Psi \rangle = -\Re\langle\hat H_0\Psi, \hat\rho_i\Psi\rangle.
\end{equation}
We introduce the $M\times M$ matrix $\Gram(\Psi)$ with $\Gram_{ij}(\Psi):=\langle \hat\rho_i\Psi, \hat\rho_j\Psi \rangle$, the Gram matrix of the Hilbert-space vectors $\hat\rho_i\Psi$. As such, $\Gram(\Psi)$ is positive semi-definite and since all $\hat\rho_i$ commute it is also real symmetric. In Appendix~\ref{app:sol-beta} it is proven that we can always solve Eq.~\eqref{eq:constraints-beta-simple} for the $v_j$, and that even if the solution $\vv$ is non-unique the evolution equation is well-defined. By putting the solution to Eq.~\eqref{eq:constraints-beta-simple} at every time $\tau$ into $\hat G(\tau)$ we have thus transformed Eq.~\eqref{eq:evolution} into an autonomous but non-linear evolution equation,
\begin{equation}\label{eq:evolution-nl}
    -\partial_\tau\Psi=\hat G(\Psi)\Psi,
\end{equation}
where we now explicitly denoted the dependence of the generator on the current state $\Psi\in\H$.
Appendix~\ref{app:differentiability-beta} then shows that this evolution equation always has a solution and that as a consequence $\vv(\tau)$ is differentiable, but that the solution may become non-unique when it crosses critical points, i.e., $\Psi$ with $\det\Gram(\Psi)=0$ (see Appendix~\ref{app:gft}).
This makes sure that the whole trajectory $\Psi(\tau)$ defined by Eq.~\eqref{eq:evolution-nl} is contained in the constraint set $\M_\rrho$.

\textbf{Progression of the method:}
Next we show that the procedure gets us closer to the optimizer in every step by demonstrating that $\langle \hat H_0 \rangle_\Psi$ monotonously decreases along the prescribed path.
In the case $i=0$, Eq.~\eqref{eq:constraints-beta-simple} demands
\begin{equation}
    \langle \hat H_0 \rangle_\Psi = -\sum_{j=0}^m v_j \rho_j,
\end{equation}
which by Eq.~\eqref{eq:generator} directly yields $\langle\hat G\rangle_\Psi=0$ at all times and thus
\begin{equation}
    -v_0 = \langle \hat H_0 \rangle_\Psi + \sum_{i=1}^m v_i \rho_i = \langle\hat H\rangle_\Psi.
\end{equation}
Since $\hat H$ is the Hamiltonian, $-v_0$ is thus the energy expectation value and $v_0$ acts like a chemical potential that keeps $\langle\hat G\rangle_\Psi=0$ at all times and guarantees the normalization of the wavefunction.
We also observe that
\begin{equation}
\begin{aligned}
    \partial_\tau \langle\hat H\rangle_\Psi &= -\langle\{\hat G, \hat H\}\rangle_\Psi + \langle\partial_\tau\hat H\rangle_\Psi \\
    &= -2\langle\hat H^2\rangle_\Psi + 2\langle\hat H\rangle_\Psi^2 + \sum_{i=1}^m \langle \partial_\tau (v_i \hat\rho_i)\rangle_\Psi \\
    &= -2\langle(\hat H-\langle\hat H\rangle_\Psi)^2\rangle_\Psi + \sum_{i=1}^m (\partial_\tau v_i) \rho_i.
\end{aligned}
 \label{Hderv1}
\end{equation}
Using this we have
\begin{equation}
\begin{aligned}
    \partial_\tau \langle \hat H_0 \rangle_\Psi &= \partial_\tau \langle \hat H \rangle_\Psi - \sum_{i=1}^m \partial_\tau (v_i \langle \hat\rho_i \rangle_\Psi) \\
    &= -2\langle(\hat H-\langle\hat H\rangle_\Psi)^2\rangle_\Psi = -2\Var_\Psi(\hat H),
\end{aligned}  
\label{Hderv2}
\end{equation}
where the term involving the potentials exactly canceled. The occurrence of the variance of $\hat H$ (that is always positive) first shows that $\partial_\tau \langle \hat H_0 \rangle_\Psi\leq 0$ at any time, which means $\langle \hat H_0 \rangle_\Psi$ monotonously decreases in the process. If the variance eventually approaches zero, this indicates that $\Psi$ is close to an eigenstate of $\hat H$ (at the present time step). Since then also $v_0 \approx -\langle\hat H\rangle_\Psi$, we have $\hat G\Psi = (\hat H+v_0\hat I)\Psi\approx 0$ and the procedure slows down and practically comes to a halt at some $\vv\in\R^m$ and an associated eigenstate $\Psi$ of $\hat H(\vv)$. If this $\Psi$ is also a ground state, then this means it minimizes $\langle \hat H(\vv) \rangle_\Psi=\langle \hat H_0 \rangle_\Psi + \sum_{i=1}^m v_i\langle\hat\rho_i\rangle_\Psi$ over all $\Psi\in\M$ and consequently, if the $\langle\hat\rho_i\rangle=\rho_i$ are taken fixed, it also minimizes $\langle \hat H_0 \rangle_\Psi$ over all $\Psi\in\M_\rrho$. Thus, if $\Psi$ is a \emph{ground state} of the current $\hat H$ we have found the \emph{optimal solution} of the optimization problem, or else we got stuck at an excited state (this can be temporarily or for indefinite time and we note below how to deal with such a case). But note that not every $\rrho\in\Bdom$ is representable by a \emph{pure} ground state, so it is entirely possible that even the optimal solution comes from an excited state. Such an excited state $\Psi$ of $\hat H(\vv)$ with eigenvalue $-v_0$ naturally links to \emph{stationary} points for the functional $\Psi\mapsto\langle \hat H_0 \rangle_\Psi$ on $\M_\rrho$~\cite{Goerling1999}. This is seen by perturbing $\Psi$ with an element $\Phi$ from the tangent space of $\M_\rrho$ at $\Psi$ and by noting that with the decomposition for the Hamiltonian from Eq.~\eqref{eq:generator} and the condition in Eq.~\eqref{eq:tangent} one immediately gets
\begin{equation}
\begin{aligned}
    \lim_{\epsilon\to 0} & \frac{1}{\epsilon}\left( \langle\hat H_0\rangle_{\Psi+\epsilon\Phi} - \langle\hat H_0\rangle_{\Psi} \right) = 2\Re\langle\Phi,\hat H_0\Psi\rangle \\
    &= -2v_0\Re\langle\Phi,\Psi\rangle - 2\sum_{i=1}^m v_i\Re\langle\Phi,\hat\rho_i \Psi\rangle = 0.
\end{aligned}
\end{equation}
Now, importantly, the method always converges in terms of $\langle \hat H_0 \rangle_\Psi$ (since this quantity is bounded below by $-\|\hat H_0\|$) and $\Psi$ (since the set $\{\Psi\in\H\mid\|\Psi\|=1\}$ is compact), yet it will still diverge in $\vv$ if $\rrho$ is from the boundary of $\Bdom$ and is not representable by any eigenstate. On the other hand, this shows that every $\rrho\in\interior\Bdom$ is always representable by a pure \emph{excited} state.

\textbf{Connectedness of the constraint set:} Since the procedure defines a continuous path $\Psi(\tau)$ on $\M_\rrho$, it is important to know about this set's connectedness. We write $\Psi = \sum_{k=1}^L c_k\Phi_k$ again in the basis $\{\Phi_k\}$ that diagonalizes all $\hat\rho_i$. Then, by Eq.~\eqref{eq:constraint-in-basis}, the constraints are given by
\begin{equation}\label{eq:constraints-in-basis-2}
    \rho_i = \langle \hat\rho_i \rangle_\Psi = \sum_{k=1}^L|c_k|^2\Lambda_{ki},
\end{equation}
where $\hat\rho_i\Phi_k = \Lambda_{ki}\Phi_k$ defines the $L\times m$ matrix $\Lambda_{ki}$ of eigenvalues.
This shows that in order to satisfy the constraints only the modulus of the coefficients $c_k$ is significant, whereas the phase still remains crucial for determining an optimizer for $\langle\hat H_0\rangle_\Psi$ (the deeper reason for this being that in general $[\hat H_0,\hat\rho_i]\neq 0$). Now since changing the phase from $c_k$ to any $\e^{\i\alpha_k}c_k$ is clearly following a continuous path in $\M_\rrho$, we can restrict ourselves to $c_k\geq 0$. By setting $\lambda_k=c_k^2$ the question of connectedness of $\M_\rrho$ is then equivalent to asking if the set of all $\{\lambda_k\}$ with
\begin{equation}\label{eq:constraints-lambda}
    \lambda_k \geq 0, \quad \sum_{k=1}^L\lambda_k=1, \quad \sum_{k=1}^L\Lambda_{ki}\lambda_k = \rho_i
\end{equation}
is connected. Now the last two equalities have affine subspaces of $\R^L$ as the solution sets and their intersection is clearly connected and stays connected when only positive numbers are considered. All coefficients that fulfill Eq.~\eqref{eq:constraints-in-basis-2} are then given by $c_k=\e^{\i\alpha_k}\sqrt{\lambda_k}$ with $\alpha_k\in[0,2\pi]$. We note that even though this already restricts the search space for the given constrained-optimization problem Eq.~\eqref{eq:opt-problem}, it can be still high-dimensional. We then rely on the imaginary-time evolution given by Eq.~\eqref{eq:evolution-nl} to describe a path toward the minimizer that always respects the constraints.

\begin{figure}
\centering
\begin{tikzpicture}[every loop/.style={min distance=30mm}]
    \useasboundingbox (-.2,1.2) rectangle (4.8,-1.7);
    \draw[thick] plot [smooth cycle,tension=1.1] coordinates { (0.5,0) (0,0.9) (1.5,0.9) (1,0) (1.5,-0.9) (0,-0.9)};
    \draw[red, fill=blue] (.75,1.1) circle (.05);
    \draw[red, fill=blue] (.75,-1.1) circle (.05);
    \draw (0.8,-1.5) node {$\M_{\rrho_1}$};
    \draw[thick] (4,0) to [out=-30,in=-150, loop] (8,0);
    \draw[thick] (4,0) to [out=30, in=150, loop] (8,0);
    \draw[red, fill=red] (4,0) circle (.05);
    \draw[red, fill=blue] (4,1.12) circle (.05);
    \draw[red, fill=blue] (4,-1.12) circle (.05);
    \draw (4,-1.5) node {$\M_{\rrho_2}$};
\end{tikzpicture}
\caption{Two different situations for constraint sets $\M_\rrho$. In the first case $\rrho_1$ is regular and the manifold consists of one connected component with two points (in blue) representing the two disconnected components with real coefficients. In the second case $\rrho_2$ is critical and a singularity (in red) appears. For a definition of `regular' and `critical' densities, see Appendix~\ref{app:gft}.}
\label{fig:constraint-manifolds}
\end{figure}

\textbf{Kickstarting the method when it gets stuck:}
Although different situations for the constraint set $\M_\rrho$ arise, as depicted in Figure~\ref{fig:constraint-manifolds}, we always have a connected set as described before. Yet, \emph{disconnected} components for the allowed coefficients arise as soon as we limit ourselves to only real coefficients $\{c_k\}$. This will be the case automatically when $\hat H_0$ is real, since then the evolution described by Eq.~\eqref{eq:generator} can only produce states with real coefficients $\{c_k\}$ if we start from real ones. Since the procedure defines a continuous path $\Psi(\tau)$, we then have no chance of passing over to another component with real coefficients, where the optimum could lie. So even though the method converges, we cannot guarantee that we have reached the optimizer, since the achieved eigenstate does not need to be the ground state. But in the same instance this shows that on each connected subset of $\M_\rrho$ with real coefficients there \emph{must} lie at least one $\Psi\in\M_\rrho$ that is an eigenstate of $\hat H(\vv)$ with some $\vv\in\R^m$ if $\hat H_0$ is real. 
It is a theorem (sometimes called the ``Hohenberg-Kohn theorem, Part 1'' \cite[Th.~7]{penz2021-Graph-DFT}) that if two Hamiltonians, which differ only in the external potential, share a ground-state density $\rrho$ then they also share the corresponding ground state. Conversely, it follows that if the expectation value of $\hat H_0$ differs for two converged states belonging to different disjoint components of $\M_\rrho$, then those states cannot agree and thus they cannot be both ground states; one of them must be an excited state.
Whenever we get stuck at an eigenstate, which can also happen if we do not limit ourselves to real coefficients, we need to find a way to kickstart the procedure again in order to eventually reach the correct optimum. The strategy will be to discontinuously change $\Psi\mapsto\Psi'$ such that the constraints are still satisfied and one has $\langle \hat H_0 \rangle_{\Psi'} < \langle \hat H_0 \rangle_\Psi$. Then the imaginary-time evolution is resumed for $\Psi'$.
The suggested scheme for including discontinuous jumps into the method is to replace all $c_k$ by $c_k' = \e^{\i\alpha_k}c_k$ with a random $\alpha_k\in[0,2\pi]$. Then by Eq.~\eqref{eq:constraints-in-basis-2} the $\Psi' = \sum_{k=1}^L c_k'\Phi_k$ still fulfills the same constraints while it could simultaneously lower the expectation value of $\hat H_0$.
How exactly this can be implemented into an algorithm will be discussed in Section~\ref{sec:algo}.

\section{Algorithm and examples} 
\label{sec:algo}

Here we describe an algorithm that implements the described constrained-optimization procedure and that was also implemented for the purpose of testing the procedure. 

We first give a brief description. We always work in the basis $\{\Phi_k\}$ of $\H$ that simultaneously diagonalizes all $\hat B_i$. Then an initial state $\Psi_0=\Psi(\tau=0)\in\M_\rrho$ (equivalent to $\langle\hat\rho_i\rangle_{\Psi_0}=\rho_i$) is selected by solving Eq.~\eqref{eq:constraints-lambda} for $\lambda_k$ by standard linear optimization. We then have the freedom of choosing the phase of the coefficients $c_k=\e^{\i\alpha_k} \sqrt{\lambda_k}$ of the initial state. In the numerical experiments we either choose $\alpha_k=0$ (which can create a bias in an otherwise symmetrical situation, see Figure~\ref{fig:square-a}) or a random value. Then starts the time-step iteration.
We want to use the Crank--Nicolson scheme~\cite{press-book} for imaginary-time propagation, which is known to be numerically stable for diffusion equations~\cite{Oishi2014}, so we need the $\vv$ from Eq.~\eqref{eq:constraints-beta-simple} at a half step $\tau + \Delta\tau/2$. To this end, a subiteration is introduced that solves for $\vv(\tau + \Delta\tau/2)$ self-consistently starting from $\Psi(\tau)$. We set $\Psi_\text{sc}=\Psi(\tau)$, solve for $\vv$ from this wavefunction, and evolve $\Psi(\tau)$ with half a Euler step to get a new $\Psi'_\text{sc}$,
\begin{equation}
    \Psi'_\text{sc} = \left( \id - \frac{\Delta\tau}{2}\hat G(\Psi_\text{sc})\right)\Psi(\tau).
\end{equation}
This is repeated until the wavefunction converges to the desired accuracy; typically only a handful of subiterations are needed. We then have $\vv(\tau + \Delta\tau/2)$ at a half step that yields the generator $\hat G(\Psi_\text{sc})$ that can further be used for the implicit Crank--Nicolson scheme, where one solves the linear equation
\begin{equation}
    \left( \id + \frac{\Delta\tau}{2}\hat G(\Psi_\text{sc})\right)\Psi(\tau+\Delta\tau) = \left( \id - \frac{\Delta\tau}{2}\hat G(\Psi_\text{sc})\right)\Psi(\tau)
\end{equation}
for the propagated $\Psi(\tau+\Delta\tau)$.
After every such main iteration step, we can then test if the constraints $\langle\hat\rho_i\rangle_{\Psi(\tau)}=\rho_i$ are still obeyed and if $\langle\hat G\rangle_{\Psi(\tau)}=0$ as the theory predicts. For a successfully propagated state we can then check convergence in $\langle\hat H_0\rangle_\Psi$, i.e., see if the distance between the values of $\langle\hat H_0\rangle_{\Psi(\tau)}$ at two successive time steps is sufficiently small. If so, the procedure is terminated, or else we proceed with the next time step.
The main iteration runs much longer, with our choice of time-step length and accuracy typically a few thousand iterations where needed for convergence.
The whole procedure is detailed in pseudocode in Algorithm~\ref{alg}.

\begin{algorithm}[H]
\caption{Constrained search in imaginary-time algorithm.}\label{alg}
\begin{algorithmic}
\State $\Psi \gets$ initial state with $\Psi\in\M_\rrho$
\For{$\texttt{iter\_main} = 0\ldots\texttt{MAX\_ITER\_MAIN}$}
    \State $\Psi_\text{sc} \gets \Psi$
    \For{$\texttt{iter\_sc} = 0\ldots\texttt{MAX\_ITER\_SC}$}
        \State $\Psi_\text{sc,prev} \gets \Psi_\text{sc}$
        \State $\hat G \gets$ solve Eq.~\eqref{eq:constraints-beta-simple} with $\Psi_\text{sc}$ and construct generator
        \State $\Psi_\text{sc} \gets$ half time-step evolution with $\hat G$ on $\Psi$ (Euler method)
        \If{$\|\Psi_\text{sc}-\Psi_\text{sc,prev}\|$ is small}
            \State \textbf{break}
            \Comment{self-consistent subiteration converged}
        \EndIf
    \EndFor
    \State $E_\text{prev} \gets \langle\hat H_0\rangle_\Psi$
    \State $\Psi \gets$ full time-step evolution with $\hat G$ on $\Psi$ (Crank--Nicolson)
    \If{$\sum_{i=1}^m(\langle\hat\rho_i\rangle_\Psi-\rho_i)^2$ is not small}
    \Comment{test constraints}
        \State \textbf{error}
    \EndIf
    \If{$|\langle\hat G\rangle_\Psi|$ is not small}
    \Comment{test $\langle\hat G\rangle_\Psi=0$ condition}
        \State \textbf{error}
    \EndIf
    \If{$|\langle\hat H_0\rangle_\Psi-E_\text{prev}|$ is small}
        \State \textbf{break}
        \Comment{main iteration converged}
    \EndIf
    \If{do random-phase jump}
        \State $\Psi' \gets$ random-phase jump of $\Psi$
        \If{$\langle\hat H_0\rangle_{\Psi'} < \langle\hat H_0\rangle_\Psi$}
            \State $\Psi \gets \Psi'$
        \EndIf
    \EndIf
\EndFor
\end{algorithmic}
\end{algorithm}

Contrary to what was proposed in Section~\ref{sec:opt-method} for kickstarting the method when it gets stuck at an excited state, in the implemented algorithm the phases of the coefficients of $\Psi$ in the $\{\Phi_k\}$ basis get chosen randomly at the end of \emph{every} iteration to produce a new state $\Psi'$. If $\Psi'$ gives a lower expectation value of $\hat H_0$, it is kept as the current state, or else the procedure continues along a continuous path with $\Psi$. This has the benefit that it is possible that already in the beginning of the iteration, long before it even gets stuck in the described way, the objective $\langle \hat H_0 \rangle_\Psi$ can be lowered to $\langle \hat H_0 \rangle_{\Psi'}$ and the procedure then continues from there, thus speeding up overall convergence.

\begin{figure*}[ht]
    \begin{subfigure}{.49\textwidth}
        \centering
        \includegraphics[width=1\columnwidth]{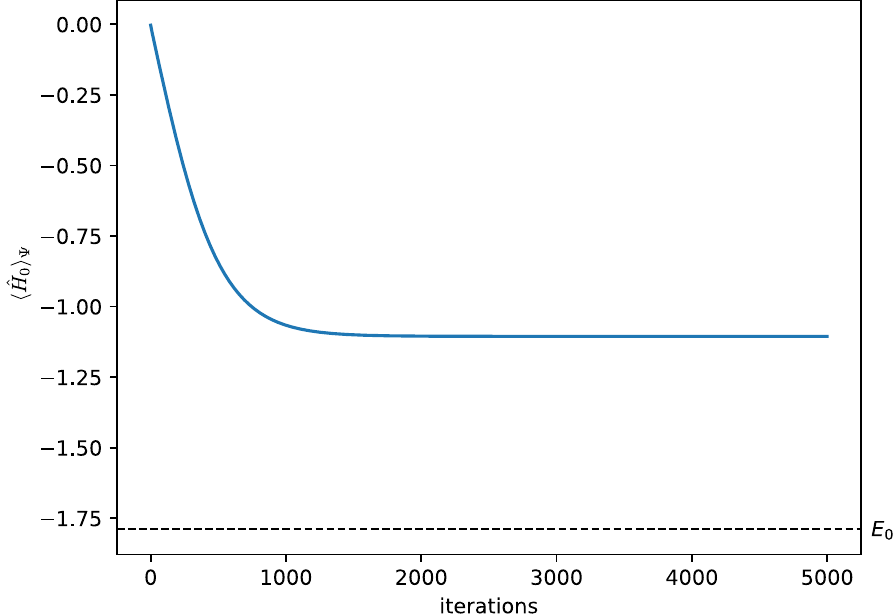}
        \caption{}
    \end{subfigure}
    \hfill
    \begin{subfigure}{.49\textwidth}
        \centering
        \includegraphics[width=1\columnwidth]{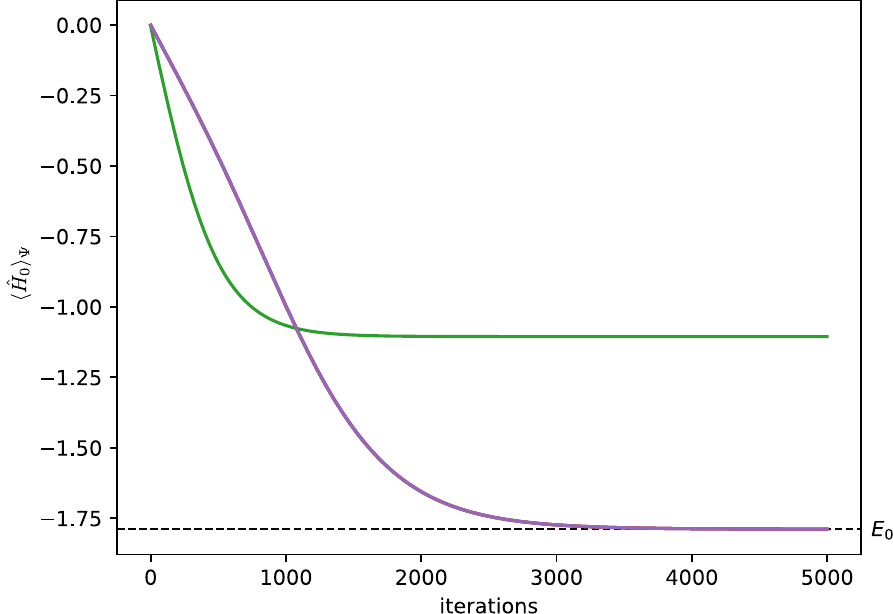}
        \caption{}
    \end{subfigure}
    \begin{subfigure}{.49\textwidth}
        \centering
        \includegraphics[width=1\columnwidth]{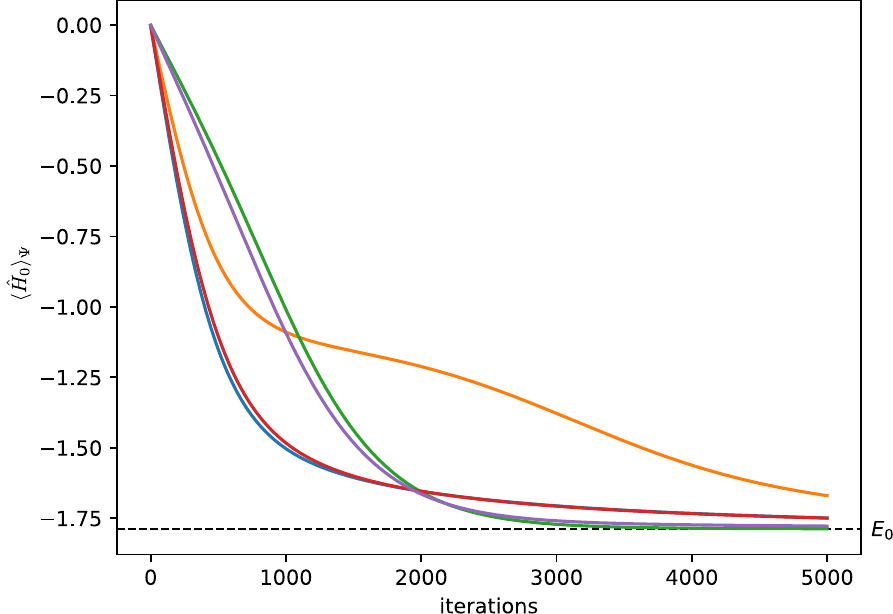}
        \caption{}
    \end{subfigure}
    \hfill
    \begin{subfigure}{.49\textwidth}
        \centering
        \includegraphics[width=1\columnwidth]{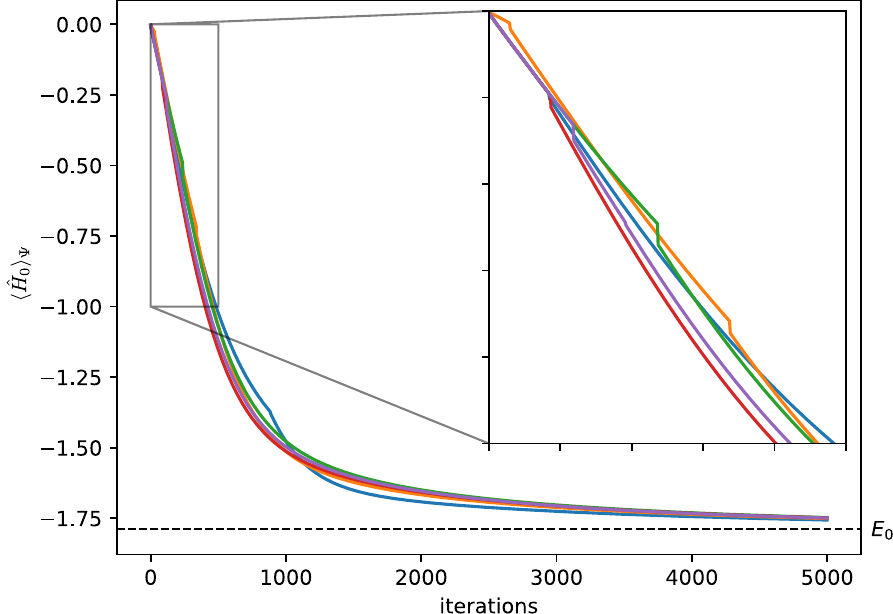}
        \caption{}
    \end{subfigure}
    \caption{Convergence behavior with different random-phase choices for the initial state and for the discontinuous jumps, tested for two particles on a square lattice. For the potential $(1,1,0,0)$ the corresponding ground-state density is held fixed during the evolution. Panel (a) shows the method without any jumps, (b) without jumps but with random signs for the initial-state coefficients, (c) without jumps but with random phases for the initial-state coefficients, and (d) with random-phase jumps during the procedure.}
    \label{fig:convergence}
\end{figure*}

\begin{figure*}[ht]
    \begin{subfigure}{.49\textwidth}
        \centering
        \includegraphics[width=1\columnwidth]{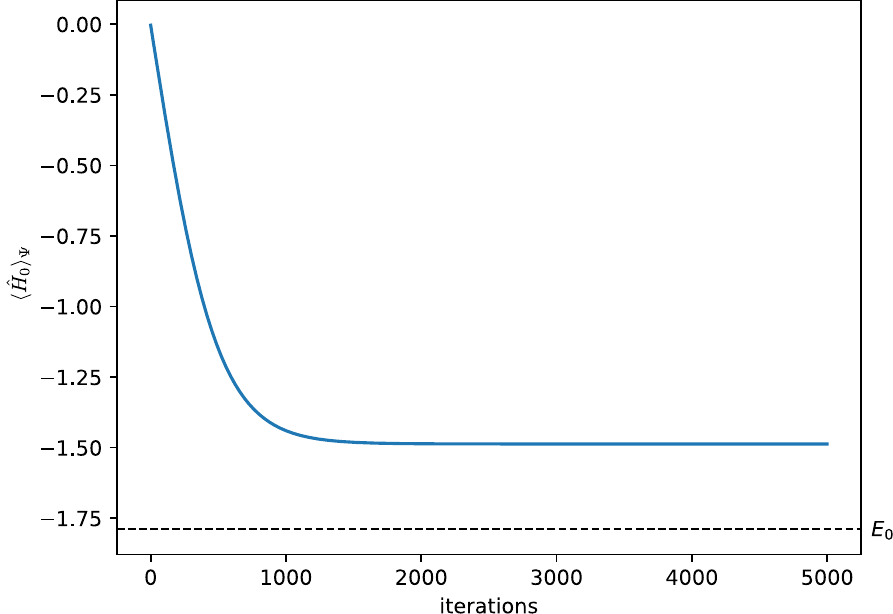}
        \caption{}
    \end{subfigure}
    \hfill
    \begin{subfigure}{.49\textwidth}
        \centering
        \includegraphics[width=1\columnwidth]{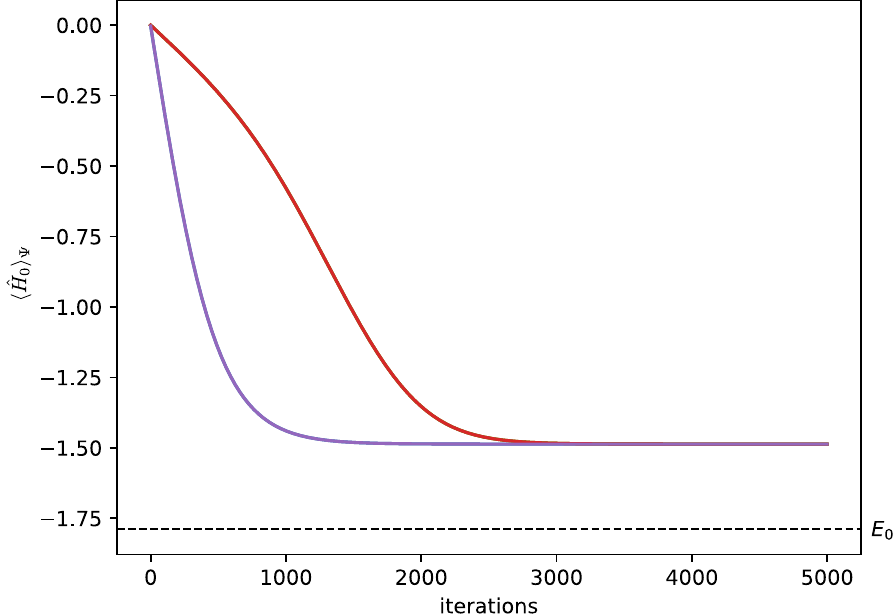}
        \caption{}
        \label{fig:convergence_deg-b}
    \end{subfigure}
    \begin{subfigure}{.49\textwidth}
        \centering
        \includegraphics[width=1\columnwidth]{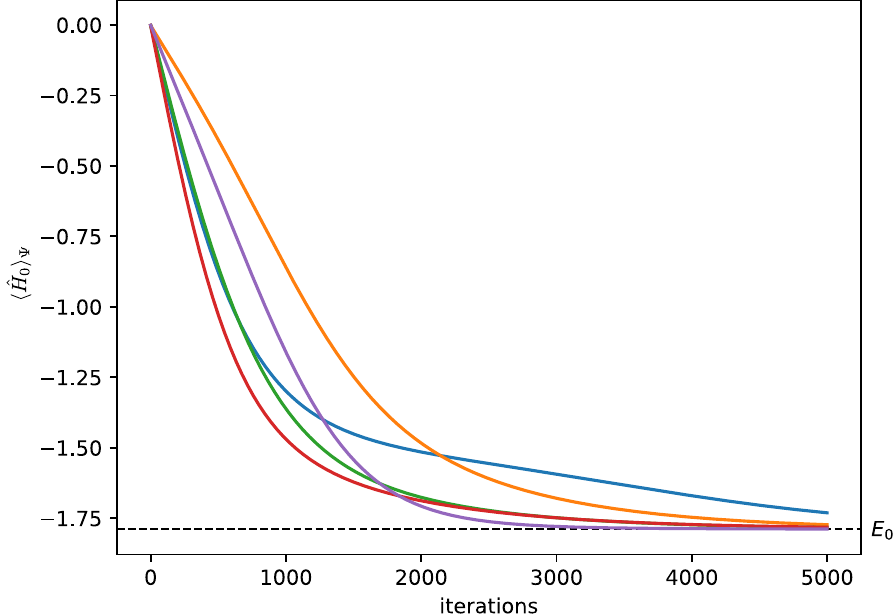}
        \caption{}
    \end{subfigure}
    \hfill
    \begin{subfigure}{.49\textwidth}
        \centering
        \includegraphics[width=1\columnwidth]{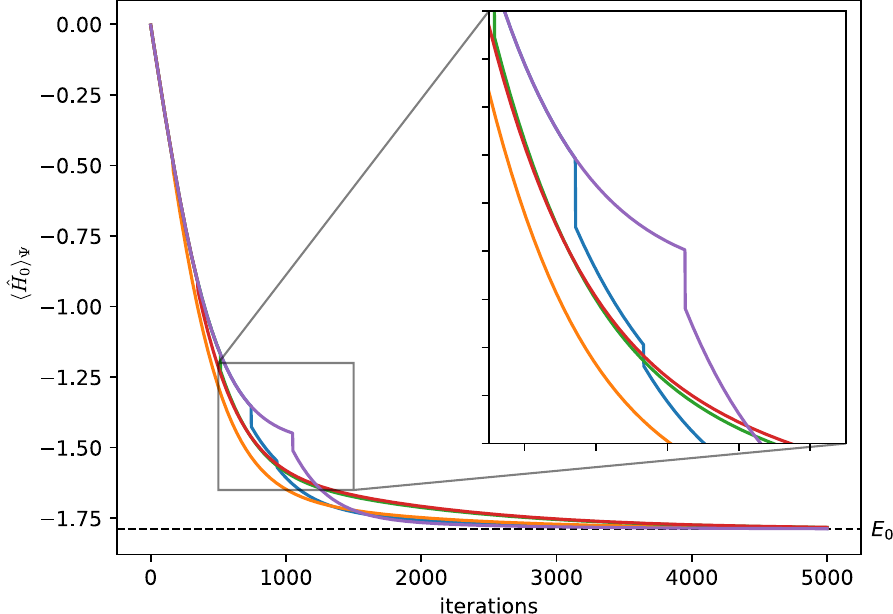}
        \caption{}
    \end{subfigure}
    \caption{Same as Figure~\ref{fig:convergence} but with the potential $(1,0,-1,0)$, which leads to two-fold degeneracy in the ground state, and a density intentionally chosen from within the degeneracy region. This means the ground state is complex and thus having $\hat G(\tau)$ real at all times and with only random signs in the initial state, it can never converge to the correct optimum in panel (b).}
    \label{fig:convergence_deg}
\end{figure*}

In order to study the newly developed optimization method, especially with respect to different phase choices and how they affect the convergence of the procedure, we apply the method to the evaluation of $\tilde F(\rrho)$ in small lattice systems with multiple fermionic particles, which were considered before in a similar context~\cite{penz2021-Graph-DFT,penz2023geometry}.
First, we take a square lattice with two particles, a problem meticulously studied before~\cite{penz2021-Graph-DFT}, and plot the convergence behavior of the procedure for two different densities, one from a ground state without degeneracy and one from within a degeneracy region (the set of all densities $\bm{\rho}=(\langle\hat\rho_1\rangle_{\Psi},\ldots,\langle\hat\rho_m\rangle_{\Psi})$ for all possible ground states $\Psi$). The results are shown in Figures~\ref{fig:convergence} and \ref{fig:convergence_deg}, where for each density we plot $\tau\mapsto\langle\hat H_0\rangle_{\Psi(\tau)}$ for different phase choices in the initial state and also allowing discontinuous jumps $\Psi\mapsto\Psi'$ as described before. We notice that without any random-phase choices, the method can easily get stuck at an excited state, but a random-phase choice for the initial state or random-phase jumps during the procedure help to converge to the desired $\langle\hat H_0\rangle_{\Psi(\tau)} \to \tilde F(\bm{\rho})$ as $\tau\to\infty$.

\begin{figure*}[ht]
    \begin{subfigure}{.33\textwidth}
        \centering
        \includegraphics[width=1\columnwidth]{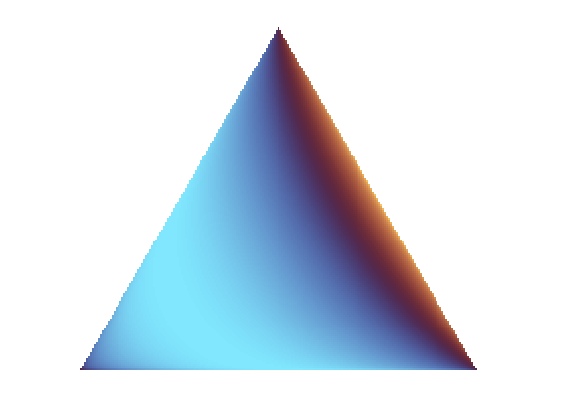}
        \caption{}
        \label{fig:triangle-a}
    \end{subfigure}
    \hfill
    \begin{subfigure}{.33\textwidth}
        \centering
        \includegraphics[width=1\columnwidth]{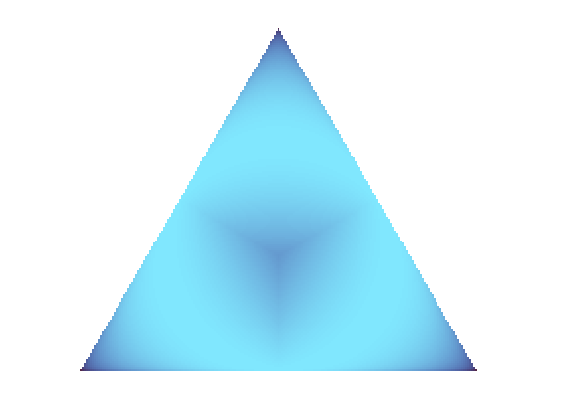}
        \caption{}
        \label{fig:triangle-b}
    \end{subfigure}
    \hfill
    \begin{subfigure}{.33\textwidth}
        \centering
        \includegraphics[width=1\columnwidth]{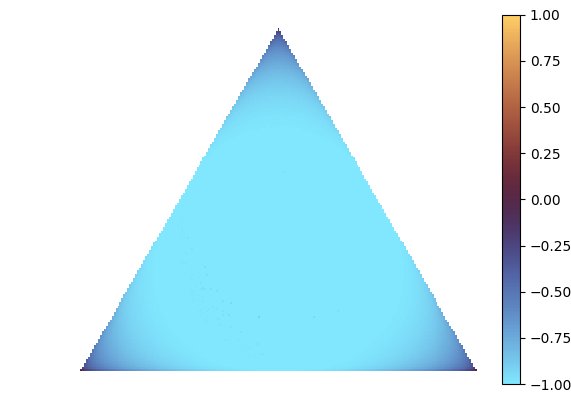}
        \caption{}
        \label{fig:triangle-c}
    \end{subfigure}
    \caption{The functional $\tilde F(\bm{\rho})$ as computed with the described procedure on a triangle lattice with two particles. Panel (a) is without any jumps, (b) with only random-sign jumps in the coefficients, and (c), which then gives the correct convex  functional, with random-phase jumps. In panel (c) the incircle of the triangle is entirely flat: this is the degeneracy region belonging to the zero potential~\cite[Sec.~VI.C]{penz2021-Graph-DFT}.}
    \label{fig:triangle}
\end{figure*}

\begin{figure}[ht]
    \begin{subfigure}{.75\columnwidth}
        \centering
        \includegraphics[width=1\columnwidth]{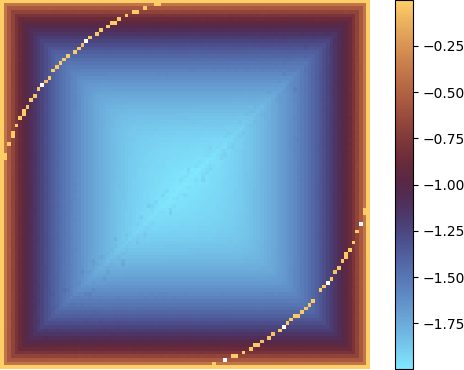}
        \caption{}
        \label{fig:square-a}
    \end{subfigure}
    \begin{subfigure}{.75\columnwidth}
        \centering
        \includegraphics[width=1\columnwidth]{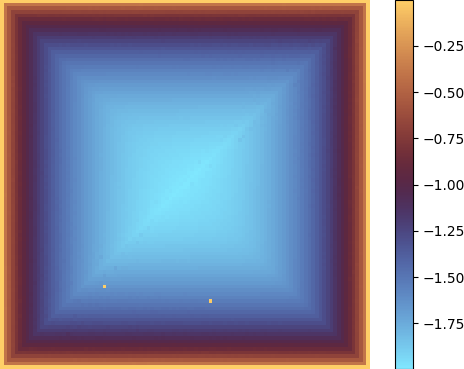}
        \caption{}
        \label{fig:square-b}
    \end{subfigure}
    \caption{The functional $\tilde F(\bm{\rho})$ on the central plane of the octahedral density domain of a square lattice with two particles as computed with the described procedure using random-phase jumps. Panel (a) is without a random-phase choice for the initial state, panel (b) is with random phases.}
    \label{fig:square}
\end{figure}

Next, in Figures~\ref{fig:triangle} and \ref{fig:square}, we give heat plots of $\tilde F(\bm{\rho})$ for all densities $\bm{\rho}\in\Bdom$ for symmetric triangle and square lattices with two particles each, so $M=3$ and 4, $N=2$, where details can again be found elsewhere~\cite{penz2021-Graph-DFT}.
In both cases we know that the functional $\tilde F(\bm{\rho})$ must be convex, since only a density that is \emph{not} representable by a pure ground state makes $F(\bm{\rho}) < \tilde F(\bm{\rho})$ and leads to $\tilde F(\bm{\rho})$ non-convex. But this can only happen with at least three-fold degeneracy~\cite{Levy1982,Lieb1983}, while the chosen examples feature two-fold degeneracy at most. So, the assured convexity of the functional $\tilde F(\bm{\rho})$ can serve as a test for our procedure. In the case of the triangle, the vertices of the density domain $\Bdom$ have occupancy $(1,1,0)$, $(1,0,1)$, and $(0,1,1)$. First, in Figure~\ref{fig:triangle-a} we always start with initial phase $\alpha_k=0$ and see that the procedure gets stuck \emph{at} excited states if no random-phase jumps are included. If those jumps are limited to random-sign switches, then the functional is correct outside the circular degeneracy region, while inside the degeneracy region we see the same effect as described in Figure~\ref{fig:convergence_deg-b} appearing: The sought-after ground state inside the degeneracy region is complex~\cite[Sec.~III]{penz2023geometry}, but a complex state can never be attained with $\hat G(\tau)$ real at all times, a real initial state, and by only switching signs in the procedure. Yet, it is still possible to reach a \emph{real excited} state with the desired density $\bm{\rho}$ for some choice of external potential. The resulting, nicely symmetric shape is that of Figure~\ref{fig:triangle-b}. Finally, by doing random-phase jumps along the iteration, we reach the correct convex functional $\tilde F(\bm{\rho})$ in Figure~\ref{fig:triangle-c}. The mentioned degeneracy region is here exactly the incircle of the triangular density domain $\Bdom$, where $\tilde F(\bm{\rho})$ is entirely flat. Note that even an analytical expression for this functional is known~\cite[Eq.~(96)]{penz2021-Graph-DFT}, where the situations in Figures~\ref{fig:triangle-a} and \ref{fig:triangle-b} are attained by analytic continuation from outside the circular degeneracy region in the center.

The square graph example in Figure~\ref{fig:square} shows the central plane of the octahedral density domain $\Bdom$ with occupancies $(1,1,0,0)$, $(0,1,1,0)$, $(0,0,1,1)$, and $(1,0,0,1)$ on the vertices (starting top left, continuing clockwise, also cp.~Figure~\ref{fig:octahedron}). Without a random initial phase but random-phase jumps during the process, Figure~\ref{fig:square-a} shows peculiarly arc-shaped sets of densities, occurring where the method is hard to converge (smaller time-steps are needed). By choosing a random phase
for the initial state, this can be resolved, see Figure~\ref{fig:square-b}, and only a very slight asymmetrical bias remains. Note that on the central plane the $\tilde F(\bm{\rho})$ is also known analytically from the ground-state energy~\cite[Eqs.~(48-49)]{penz2021-Graph-DFT} and that it precisely corresponds to the lower half of a Steinmetz solid~\cite[Fig.~6]{penz2023geometry}.
Note that in both examples, the triangle and the square lattice, it is also possible to evaluate $\tilde F(\bm{\rho})$ at the boundary of $\Bdom$, even though almost all densities on the boundary are not representable by ground states (only if a degeneracy region touches the boundary does the density at the boundary become representable by a ground state~\cite[Th.~9(b)]{penz2023geometry}). In such cases, the potential $\vv$ necessary to represent the density diverges, but already for moderately large $\vv$ the method has converged to the desired accuracy.

\begin{figure*}[ht]
    \begin{subfigure}{1\textwidth}
        \centering
        \includegraphics[width=1\columnwidth]{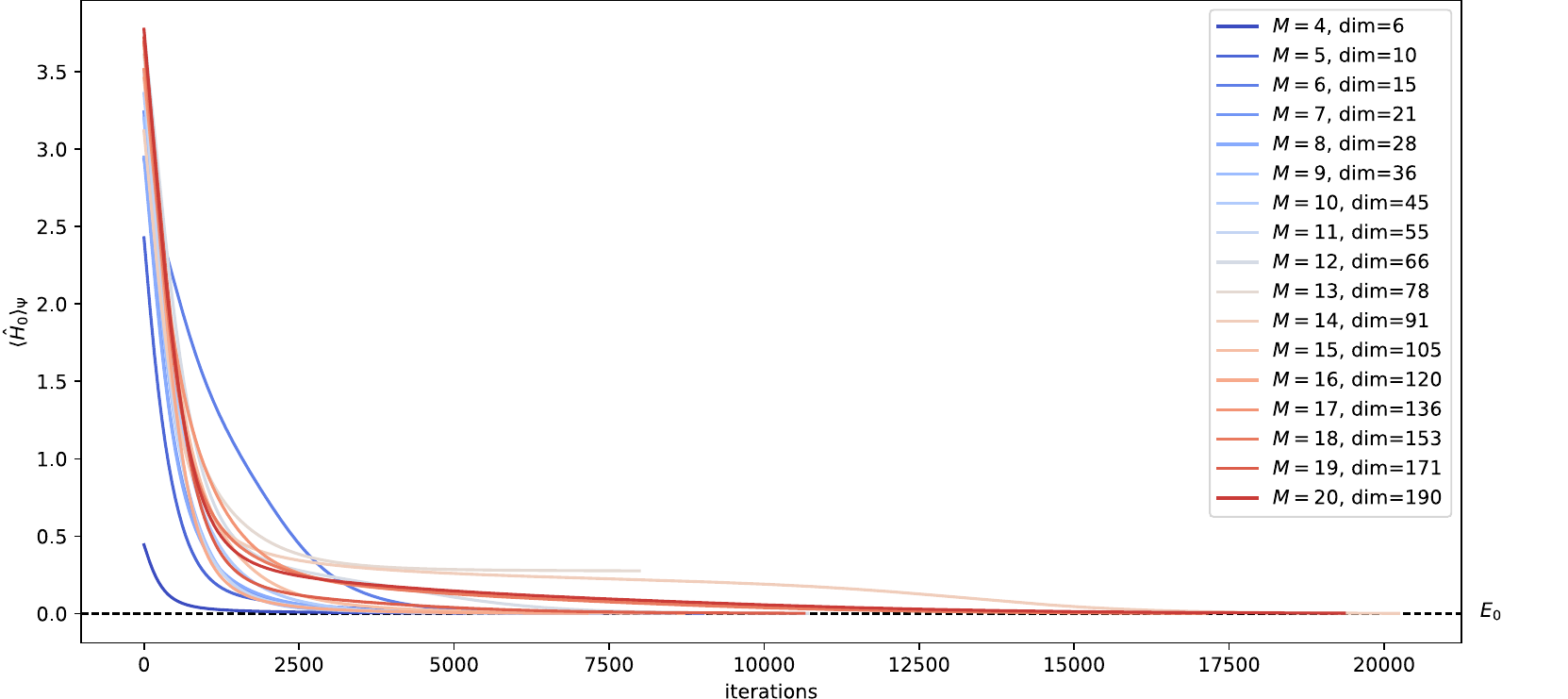}
        \caption{}
    \end{subfigure}
    \hfill
    \begin{subfigure}{1\textwidth}
        \centering
        \includegraphics[width=1\columnwidth]{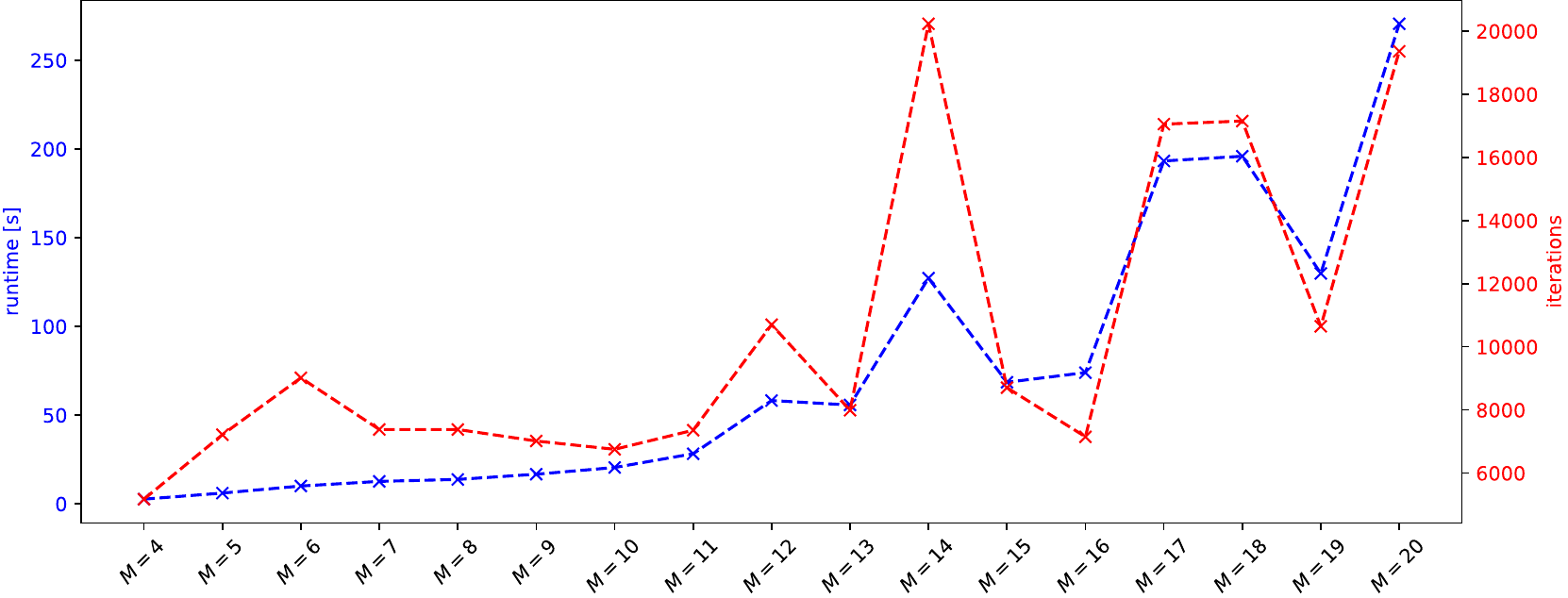}
        \caption{}
    \end{subfigure}
    \caption{The method is applied to two particles on a ring system with $M=4,\ldots,20$ lattice sites, where the ground-state density of the potential $v=(0.1,0,\ldots,0)$ is held fixed. Panel (a) shows the convergence behavior, where no random elements like random-phase choices for initial states or jumps were introduced. The method always converges to the correct value, except for $M=13$ where it gets stuck at an excited state. The dimension of the Hilbert space is shown together with the lattice size in the legend. Panel (b) shows how the computation time and main iteration number scales.}
    \label{fig:ring-scaling}
\end{figure*}

Finally, we test the method on a larger lattice. We choose a ring lattice system with variable size $M$ ranging from $4$ to $20$ sites and always take $N=2$ particles. The fixed density $\rrho$ for all runs is the ground-state density for the potential $v=(0.1,0,\ldots,0)$. In order to have a reproducible behavior, no random initial phase or discontinuous jumps were introduced. Still, the method converges to the correct value in all cases but one ($M=13$), as shown in the upper panel of Figure~\ref{fig:ring-scaling}. In the case $M=13$ we have again the situation of getting stuck at an excited state, while for $M=14$ we notice how the method slows down close to an excited state (visible as an plateau in the curve) but finally finds its way to the correct minimum. The lower panel shows the computation time (on an Intel Core i7-7700HQ CPU) and the necessary main iteration number. We see that the method scales fairly modestly with the system size, even though the Hilbert-space dimension increases significantly and can even yield faster results for larger systems in certain cases.

\section{Summary and outlook}
\label{sec:sum}

We described an optimization method for minimizing $\langle\hat H_0\rangle_\Psi$, ${\hat H_0}$ self-adjoint, over a finite-dimensional Hilbert space $\H$ under the constraints $\langle\hat\rho_i\rangle_\Psi=\rho_i$, where the $\hat\rho_i$ are the density operators (self-adjoint, linearly independent together with $\id$, and mutually commuting). While the method is entirely general, as demonstrated in Appendix~\ref{app:gft}, it has a particular relevance for finite-lattice DFT, since it yields the pure-state constrained-search functional $\tilde F(\rrho)$ if fully converged to the optimum. The method can be used to determine this functional over a domain in $\Bdom$ but also as a single-shot method for one density $\rrho\in\Bdom$. In the discussion we noted that non-uniqueness can appear in the trajectory of the procedure and, for the same reasons, also in the density-potential mapping, thus yielding counterexamples to the Hohenberg--Kohn theorem. In order to not get stuck at excited states of the constructed Hamiltonian $\hat H(\vv)$, we had to introduce discontinuous random jumps into the method. Even though the method still proceeds in a monotonous way, we cannot guarantee its convergence to the global optimum. An analysis of the resulting $\tilde F(\rrho)$ over a larger domain will usually help to decide if the correct functional has been attained. Further, although the method is aimed at yielding a minimum for the objective $\langle\hat H_0\rangle_\Psi$, it also produces a corresponding potential $\vv\in\R^m$ such that the Hamiltonian $\hat H(\vv)=\hat H_0+\sum_{i=1}^m v_i\hat\rho_i$ has a ground state (or excited state in the case of premature convergence or if the density is not pure-state representable) $\Psi$ obeying the constraints $\langle\hat\rho_i\rangle_\Psi=\rho_i$. This is equivalent to giving the subdifferential $-\vv\in\partial\tilde F(\rrho)$~\cite[Sec.~VII]{Penz2023-Review-Part-I}, so with a single run we additionally gain semilocal information about the functional. This puts the method into the context of inverse Kohn--Sham methods~\cite{Shi2021,vanleeuwen1994exchange,Schipper1997,Schipper1998,Penz2023-MY-ZMP} that aim at inverting from densities to potentials. But contrary to those methods, our procedure is not specialized to non-interacting particles and always operates on the whole many-particle Hilbert space.
Interestingly, since $\hat H(\vv)$ will always conserve the exchange symmetry of a state, this means that it does not even matter if the Hilbert space $\H$ itself is restricted to this symmetry or not, just the initial state needs to respect it. The method itself is thus the same for fermions, bosons, or qubits.
Since the current formulation of the method is limited to finite-dimensional Hilbert spaces, a first and obvious question is about the extensibility of the framework to the usual infinite-dimensional Hilbert spaces of quantum mechanics. But already the finite-dimensional setting allows for a multitude of interesting examples, with model systems such as those of Hubbard, Ising, and Dicke coming to one's mind.

Within the procedure the Gram matrix given by
\begin{equation}
    \Gram_{ij} (\Psi) = \langle \hat\rho_i \Psi, \hat\rho_j \Psi \rangle = \langle \Psi, \hat\rho_i \hat\rho_j \Psi \rangle
\end{equation}
took a special role. It is just the diagonal of the two-body reduced-density matrix that closely relates to the pair-correlation function and to the exchange-correlation hole~\cite[Sec.~2.4]{parr-yang-book}. This raises the idea that one could also put reduced-density matrix functional theory (RDMFT) into the presented framework. Yet, the corresponding $\hat\rho_{ij}$ that would give the one-body reduced-density matrix as an expectation value do not commute. This implies that we cannot find a basis for $\H$ that diagonalizes all $\hat\rho_{ij}$ simultaneously, as it is possible for $\hat\rho_{i}$, and that consequently, the problem of representability by pure states is much more complicated~\cite{altunbulak2008pauli} and demands a long list of conditions that are known as `generalized Pauli constraints'~\cite{Schilling2018}. Yet, it must be stressed that the procedure itself does not rely on the commutativity condition; we just need it to easily determine an initial state and for performing the random-phase jumps.
On the other hand, within the commutative framework used here and laid out in more detail in Appendix~\ref{app:gft}, a comprehensive characterization of representability can be established: First, every $\bm{\rho}\in\Bdom$ is representable by a pure state $\Psi\in\H$. Then, if $\bm{\rho}\in\interior\Bdom$ there is a potential $\vv\in\R^m$ with possibly degenerate ground states $\{\Psi_k\}$ such that an ensemble from these states represents $\bm{\rho}$. Although the situation that a density is only representable by a ground-state ensemble instead of a pure ground state is clearly the exception, it is known that if they occur, the set of only-ensemble ground-state representable densities has a non-zero measure~\cite[Cor.~7]{penz2023geometry}, so they are not rare either.
Finally, the convergence of the optimization method itself shows that every $\bm{\rho}\in\interior\Bdom$ is also representable by a pure \emph{excited} state.

Summing up, apart from offering a new optimization method, this line of research also allows one to achieve a furthered understanding about the mathematical basics of DFT and related theories and even offers new and interesting proof techniques. These techniques should be especially useful to study time-dependent DFT on a lattice and an improved formulation of its basic theorems~\cite{li2008time,Farzanehpour2012} is already underway.
As a final remark we mention that imaginary-time propagation techniques have received considerable attention~\cite{McArdleetal2019,Mottaetal2020,Turroetal2022,Sokolovetal2023} in quantum-computing applications.
Nevertheless, in quantum computing, the term ``constrained optimization'' usually refers to constraining quantum evolution to a subspace~\cite{Herman2023}, instead of enforcing expectation-value constraints through the continuous adaptation of the potential. Our method is thus critically different, and it is imaginable that this constrained search in imaginary time can be adapted as a quantum (or rather hybrid quantum-classical) algorithm.

\section*{Data availability}
The method was implemented as the extension \texttt{coptimize} of the self-developed Python package \texttt{qmodel}~\cite{qmodel-coptimize}. The repository includes the scripts for all the displayed plots as well as further examples.

\begin{acknowledgments}
M.P.\ thanks Christian Schilling and his group for an invitation and discussions that helped to refine some of the statements about generalized functional theories. 
R.v.L.\ acknowledges support from the Suomen Akatemia (Finnish Academy) under Project No.~356906.
\end{acknowledgments}

\appendix

\section{Generalized functional theories}
\label{app:gft}

The method developed here can be generalized such that it is also applicable beyond the discussed version of lattice DFT. Yet, a precise formulation makes it necessary to also find a fitting and generalizable framework for DFT itself, which will be the content of this section. This framework, although simplified in that it builds on a finite-dimensional Hilbert space, is complete in the sense that it contains a rigorous Hohenberg--Kohn theorem and a full solution to the infamous $v$-representability problem. It is also not limited to any particular choice of density variables but replaces them by a vector $\bb\in\R^m$ that contains just the expectation values $b_i=\langle\hat B_i\rangle_\Psi$ of self-adjoint operators $\hat B_i$. In this way, it automatically encompasses finite-lattice density-functional theory~\cite{penz2021-Graph-DFT} together with its variants that describe spin densities, currents, etc. Such generalized theories have been suggested before in the literature~\cite{Schoenhammer1995,higuchi2004arbitrary,higuchi2004arbitrary2}, and amazingly, they even predate DFT itself in a closely related form~\cite{DeDominicis1963} and served as an inspiration for its formulation~\cite{Hohenberg1990}.
However, this was previously not done in the form of a complete mathematical framework.

\textbf{Setting and general assumptions:} We take a finite-dimensional complex Hilbert space $\H$ of dimension $L=\dim\H$ with inner product $\langle\cdot,\cdot\rangle$ that is antilinear in the first component. On it, we have self-adjoint operators $\hat A, \hat B_1, \ldots, \hat B_m$ such that the set $\{\hat B_0=\id,\hat B_1,\ldots,\hat B_m\}$ (including the identity operator $\id$) is linearly independent (meaning that no operator can be written as the linear combination of the others and $L\geq m+1$) and all $\hat B_i$ mutually commute. We take $\hat A$ as the \emph{universal} part of the system's Hamiltonian and the $\hat B_i$ as the observables coupling to a given generalized potential $\bbeta\in\R^m$. This yields the Hamiltonian
\begin{equation}\label{eq:Ham}
    \hat H(\bbeta) := \hat A + \sum_{i=1}^m\beta_i\hat B_i.
\end{equation}
Since we will always be concerned with normalized states, we define the compact state manifold $\M := \{\Psi\in\H\mid\|\Psi\|=1\}$.
The ground-state problem has then the variational formulation
\begin{equation}\label{eq:gs-problem}
    E(\bbeta) := \min_{\Psi\in\M} \langle\hat H(\bbeta)\rangle_\Psi,
\end{equation}
and the primary aim of the functional theory is to substantially decrease the complexity of this optimization problem by reducing the search space.
To achieve this, we define the vector-valued map from states in $\M$ to the expectation values of the general observables
\begin{align}
    &\gmap(\Psi):=(\langle\hat B_1\rangle_\Psi,\ldots,\langle\hat B_m\rangle_\Psi)\in\R^m.
\end{align}
This allows us to rewrite the expectation value of $\hat H(\bbeta)$ compactly as $\langle\hat H(\bbeta)\rangle_\Psi = \langle\hat A\rangle_\Psi + \bbeta\cdot\gmap(\Psi)$.
The universal part can then be treated in a separate variational problem. We define the constraint set $\M_\bb := \{ \Psi\in\M \mid  \gmap(\Psi)=\bb\} = \gmap^{-1}(\bb)$ and the pure-state constrained-search functional
\begin{equation}\label{eq:F-tilde-def}
    \tilde F(\bb):=\left\{ 
	\begin{array}{ll}
		\min\limits_{\Psi\in\M_\bb}\langle\hat A\rangle_\Psi & \text{if}\;\bb\in\Bdom\\
		\infty & \text{else.}
	\end{array}	
	\right.
\end{equation}
Here, the effective domain $\Bdom\subseteq\R^m$ of $\tilde F$ (later shown to be compact) amounts to the image of $\M$ under $\gmap$. Equation~\eqref{eq:F-tilde-def} can be further treated as a Lagrange-multiplier problem like in \citet{Bakkestuen2025}, but will not follow this connection further on this occasion. The ground-state problem Eq.~\eqref{eq:gs-problem} can now be reformulated with the universal functional and a substantially reduced search space $\Bdom$,
\begin{equation}\label{eq:E-variation}
    E(\bbeta) = \min_{\bb\in\Bdom} \{\tilde F(\bb) + \bbeta\cdot\bb \}.
\end{equation}
Since the energy functional $E(\bbeta)$ is concave by its definition in Eq.~\eqref{eq:gs-problem}, the functional theory can be transformed into a fully convex form. This is achieved via the Legendre--Fenchel transform of $E(\bbeta)$ that defines the \emph{convex} universal functional~\cite{Lieb1983}
\begin{equation}\label{eq:F-def}
    F(\bb):=\sup_{\bbeta\in\R^m}\{E(\bbeta)-\bbeta\cdot\bb\}.
\end{equation}
It can be shown that $F(\bb)$ is the same as Eq.~\eqref{eq:F-tilde-def} if the constrained search is extended to ensemble states~\cite{Lieb1983} and as such it is the convex hull of $\tilde F(\bb)$~\cite[Prop.~18]{penz2021-Graph-DFT}. Thus, $F(\bb)$ has the same proper domain $\Bdom$ and $F(\bb)\leq \tilde F(\bb)$. It should be noted that the supremum above is not necessarily a maximum since now the search space is not compact. We will later give conditions when it is indeed a maximum and an optimizer $\bbeta\in\R^m$ always exists.

\textbf{Representability by pure states:} We already defined $\Bdom$ as the set of all values $\bb=\gmap(\Psi)\in\R^m$ that can be achieved by states $\Psi\in\M$. But in order to work with this set as the new search space for the ground-state problem, we need to find a more explicit description for it that also yields a constructive scheme to determine a $\Psi\in\M_\bb$. Since the $\hat B_i$ (including $i=0$) all commute, there is an orthonormal basis $\{\Phi_k\}$ of $\H$ in which all those operators are simultaneously diagonal,
\begin{equation}
	\hat B_i\Phi_k = \Lambda_{ki}\Phi_k, \quad \gmap(\Phi_k)=(\Lambda_{k1},\ldots,\Lambda_{km}),
\end{equation}
and $\Lambda_{k0}=1$. This defines a real $L\times (m+1)$ matrix $\Lambda_{ki}$ of eigenvalues. Since the $\hat B_i$ were further assumed all linearly independent, also the $m+1$ columns of the matrix $\Lambda_{ki}$ are, which means the matrix has rank $m+1$. Every $\Psi\in\M$ can now be written $\Psi=\sum_{k=1}^L c_k\Phi_k$, $c_k\in\C$, $\sum_{k=1}^L|c_k|^2=1$. The expectation value of $\hat B_i$ under this state is then
\begin{equation}\label{eq:constraint-in-basis}
\begin{aligned}
    \langle\hat B_i\rangle_\Psi &= \sum_{k,l=1}^Lc_k^*c_l\langle\Phi_k,\hat B_i\Phi_l\rangle \\&= \sum_{k,l=1}^Lc_k^*c_l\Lambda_{li}\langle\Phi_k,\Phi_l\rangle = \sum_{k=1}^L|c_k|^2\Lambda_{ki}.
\end{aligned}
\end{equation}
Writing $\lambda_k=|c_k|^2$ this is equivalent to
\begin{equation}\label{eq:pure-rep}
	\gmap(\Psi) = \sum_{k=1}^L\lambda_k\gmap(\Phi_k), \quad \lambda_k\geq 0,\sum_{k=1}^L\lambda_k=1.
\end{equation}
Thus, $\Bdom$ is the convex hull of $\{\gmap(\Phi_k)\}$ and the $\gmap(\Phi_k)$ form the vertices of the polyhedron $\Bdom$. This also shows that $\Bdom$ is closed and consequently it is compact as a subset of $\R^m$. Given any $\bb\in\Bdom$, we can find a $\Psi\in\M_\bb$ by determining a (in general non-unique) set $\{\lambda_k\}$ that solves Eq.~\eqref{eq:pure-rep} and then taking $c_k=\sqrt{\lambda_k}$ (or with any other phase choice).

\textbf{Representability by ensemble ground states:} Yet, the above form of representability by pure states alone is not enough for our purpose, since we will later see that in the optimization method we aim at a $\Psi\in\M$ that is the \emph{ground state} of some Hamiltonian $\hat H(\bbeta)$. To be more specific, note that from Eq.~\eqref{eq:F-tilde-def} it follows that $F(\bb)\leq\tilde F(\bb)\leq\|\hat A\|<\infty$ on $\Bdom$ and thus by some standard results from convex analysis (see for example \citet[Th.~2.14, Prop.~2.36, and Prop.~2.33]{Barbu-Precupanu}) it holds that for every $\bb$ from the \emph{interior} of $\Bdom$ there is a $\bbeta\in\R^m$ such that $E(\bbeta)=F(\bb)+\bbeta\cdot\bb$. In other words, for all $\bb\in\interior\Bdom$ the supremum in Eq.~\eqref{eq:F-def} is a maximum where an optimizer $\bbeta\in\R^m$ can be found. But since the functional involved is given by constrained search over \emph{ensemble} states, the corresponding ground state of $\hat H(\bbeta)$ that maps to $\bb$ might be an ensemble state itself.
\Citet{CCR1985} used a completely different method for showing the representability by ensemble ground states for infinite lattice settings, and we expect that this method can be also used in the setting employed here. It should be added that in exceptional situations linked to degeneracies~\cite{penz2023geometry}, the representability by ground states also holds for $\bb$ on the boundary of $\Bdom$. The fact that one actually needs to rely on ensemble states is shown with examples where a $\bb\in\interior\Bdom$ is not representable by pure ground states~\cite{penz2021-Graph-DFT}. Finally, the optimization method developed here shows that still every $\bb\in\interior\Bdom$ is representable by a \emph{pure excited state}. This fits to recent findings in the Hubbard dimer~\cite{Giarrusso2023} and Dicke model~\cite{Bakkestuen2025}.

\textbf{Unique representability (regular and critical values, Hohenberg--Kohn property):} The last element of our generalized functional theory is to ask if, or when, the above form of ground-state representability with a $\bbeta\in\R^m$ is unique. In DFT this is the content of the Hohenberg--Kohn theorem~\cite{hohenberg-kohn1964,Garrigue2018HK}. Since in what follows we only consider pure states, a more general ensemble-state formulation will be delayed to later work.
A $\Psi\in\H$ is called \emph{regular point} if all the $\hat B_i\Psi$, $i=0,\ldots,m$, are linearly independent, or else it is called a \emph{critical point}. The set of all critical points is denoted $\crit\subseteq\H$. The image of $\crit\cap\M$ under $\gmap$ gives the \emph{critical values} $\Bcrit\subseteq\Bdom$; a $\bb\in\Bdom\setminus\Bcrit$ is then called a \emph{regular value}. In other words, a $\bb\in\Bdom$ is called regular if for all $\Psi\in\M_\bb$ all the $\hat B_i\Psi$, $i=0,\ldots,m$, are linearly independent, otherwise they are called critical. This explains why the $\hat B_i$ have been chosen linearly independent in the first place, because else no $\bb$ can ever be regular.
An alternative characterization of the critical points can be given with the help of the Gram matrix $\Gram(\Psi)$ defined by $\Gram_{ij}(\Psi):=\langle \hat{B}_i \Psi, \hat{B}_j \Psi \rangle$, $i,j=0,\ldots,m$. This matrix will have an important role later in the formulation of the optimization method. A very similar matrix and argument is used by \citet{Xu2022} in their discussion of the Hohenberg--Kohn theorem, albeit without the commutativity condition on $\hat B_i$.
Since the vectors $\hat{B}_i \Psi$ are linearly independent if and only if $\det\Gram(\Psi)=0$, the critical points $\Psi\in\crit$ are given by an algebraic condition that amounts to the vanishing of a polynomial in terms of the wavefunction coefficients. Further useful relations between the Gram matrix and the critical points are established in Appendix~\ref{app:crit}.
\\
Then the Hohenberg--Kohn theorem in the current setting is the following: If the pure ground states (degeneracy is possible) of $\hat H(\bbeta)$ and $\hat H(\bbeta')$ yield the same \emph{regular} $\bb$ then $\bbeta=\bbeta'$. Proof: For fixed $\bb$ and $\bbeta,\bbeta'$ the ground-state energies are $E(\bbeta^{(\prime)}) = \min_{\Psi\in\M_\bb} \langle \hat A \rangle_\Psi + \bbeta^{(\prime)}\cdot\bb$. Since the minimum is independent of $\bbeta^{(\prime)}$, we can take the same $\Psi$ for both Hamiltonians. Subtracting the two Schrödinger equations $\hat H(\bbeta^{(\prime)})\Psi=E(\bbeta^{(\prime)})\Psi$ gives
\begin{equation}\label{eq:HK}
\begin{aligned}
    &\sum_{i=1}^m(\beta_i-\beta'_i)\hat B_i\Psi = (\overbrace{E(\bbeta)}^{-\beta_0}-\overbrace{E(\bbeta')}^{-\beta'_0})\Psi \\
    \Longrightarrow &\sum_{i=0}^{m}(\beta_i-\beta'_i)\hat B_i\Psi=0.
\end{aligned}
\end{equation}
Since all $\hat B_i\Psi$ are linearly independent, it readily follows $\beta_i=\beta'_i$ for all $i=0,1,\ldots,m$. This concludes the proof.
The classical Sard's theorem~\cite{abraham2012manifolds} applied to $\gmap:\M\to\Bdom$ then tells us that the set of critical values $\Bcrit$ has measure zero in $\R^m$, so almost all $\bb\in\Bdom$ are regular and allow for the Hohenberg--Kohn property.
It follows that only for critical $\bb$ can counterexamples to the Hohenberg--Kohn theorem be found~\cite{penz2021-Graph-DFT}, and it is known that those $\bb\in\Bcrit$ that produce counterexamples always correspond to the intersection of so-called degeneracy regions in $\Bdom$~\cite{penz2023geometry}.\\
While we do not need the Hohenberg--Kohn property directly in the context of our work here, the notions of regular and critical points will be relevant for what follows. For a regular value $\bb\in\Bdom\setminus\Bcrit$ we also know that the constraint set $\M_\bb$ is a closed submanifold (constraint manifold) of $\M$ by the submersion theorem~\cite{abraham2012manifolds}. We note that $\M_\bb$ can consist of multiple connected components and that a tangent vector $\Phi$ to $\M_\bb$ at any $\Psi\in\M_\bb$ is defined by the condition
\begin{equation}\label{eq:tangent}
    \lim_{\epsilon\to 0}\frac{1}{\epsilon}\left( \langle\hat B_i\rangle_{\Psi+\epsilon\Phi} - \langle\hat B_i\rangle_{\Psi} \right) = 2\Re\langle\Phi,\hat B_i\Psi\rangle = 0.
\end{equation}
Note that this does not mean that the $\hat B_i\Psi$ are normal vectors to the constraint manifold $\M_\bb$ at $\Psi\in\M_\bb$ in the usual Hilbert-space sense, but they are orthogonal to $\M_\bb$ with respect to the Kähler metric that is exactly defined by $g(\Phi,\Psi)=\Re\langle\Phi,\Psi\rangle$~\cite{ashtekar1999geometrical}.
Different possible situations for the constraint manifold are depicted in Figure~\ref{fig:constraint-manifolds}.\\
These discussions of representability, regular and critical values, and the Hohenberg--Kohn property for regular $\bb\in\Bdom$ finalize this brief mathematical treatment of generalized functional theories, a setting in which we will now formulate our optimization method.
It should be noted, though, that the reduced-density-matrix functional theory is currently not contained in our formulation since the $\hat B_i$ must be mutually commuting operators.
As a final note, we want to highlight a noteworthy recent paper discussing the geometry of expectation values by \citet{Song2023} that employs very similar concepts and even features an algebraic formulation of DFT, although a closer relation to our work yet remains to be established.

\textbf{Optimization method:} We conclude this appendix with a reformulation of the developed optimization method from Section~\ref{sec:opt-method} in the generalized setting.
\begin{equation}\label{eq:opt-problem-gen}
\begin{aligned}
    &\min_{\Psi\in\H} \langle\hat A\rangle_\Psi  \quad\text{under the constraints}\\
    &\langle \hat B_0\rangle_\Psi = \|\Psi\|^2 = 1 \quad\text{and}\\
    &\langle \hat B_i\rangle_\Psi = b_i \;\;\text{for all}\;\; i=1,\ldots,m.
\end{aligned}
\end{equation}
In shorthand notation this is equivalent to
\begin{equation}
    \min_{\Psi\in\M_\bb} \langle\hat A\rangle_\Psi.
\end{equation}
As was demonstrated, the method consists in performing an imaginary-time evolution with the generator
\begin{equation}\label{eq:generator-gen}
    \hat G(\tau) = \hat H(\tau) + \beta_0(\tau)\id,\quad \hat H(\tau) = \hat A + \sum_{i=1}^m \beta_i(\tau) \hat B_i,
\end{equation}
and any initial state $\Psi_0\in\M_\bb$, which can be found by solving Eq.~\eqref{eq:pure-rep}, and where at each time the $\beta_i(\tau)\in\R$ come from the solution of
\begin{equation}\label{eq:constraints-beta-simple-gen}
     \sum_{j=0}^m \beta_j \Gram_{ij}(\Psi) = -\Re\langle\hat A\Psi, \hat B_i\Psi\rangle.
\end{equation}
In Appendix~\ref{app:sol-beta} it is proven that we can always find a solution and that even if the solution $\bbeta$ is non-unique the evolution equation is well-defined. Appendix~\ref{app:differentiability-beta} then shows that the resulting evolution equation always has a solution and that as a consequence $\bbeta(\tau)$ is differentiable, but that the solution may become non-unique when it crosses critical points. We will keep the notation of the generalized setting in the appendices that follow.

\section{Gram matrix and critical values}
\label{app:crit}

By expanding $\Psi\in\H$ in the common orthonormal eigenstates $\Phi_k$ of all the $\hat{B}_i$ as $\Psi=\sum_{k=1}^L c_k \Phi_k$, we find that the Gram matrix is equivalently given by
\begin{equation}
\Gram_{ij}(\Psi) = \sum_{k,l=1}^L c_k^* c_l \langle \hat{B}_i \Phi_k, \hat{B}_j \Phi_l \rangle = \sum_{k=1}^L |c_k |^2 \Lambda_{ki} \Lambda_{kj}.
\end{equation}
Since $\Psi\in\crit$ iff $\det\Gram(\Psi)=0$, we can already infer that the critical points are completely determined by the absolute values $|c_k|$ of the wavefunction coefficients and no phase information is needed. Now $\det\Gram(\Psi)=0$ can only happen when $\Gram(\Psi)$ has a zero eigenvalue or, equivalently, has a non-zero kernel. Since $\Gram (\Psi)$ is real and symmetric we can always choose an orthonormal set of real eigenvectors.
Let us therefore consider
a real vector $\uu \in \ker\Gram(\Psi)$. Then we have
\begin{equation}
0 = \sum_{i,j=0}^{m} u_i \Gram_{ij}(\Psi) u_j = \sum_{k=1}^L \alpha_k^2 |c_k|^2 ,
\label{kernel_eqn}
\end{equation}
where we defined
\begin{equation}
\alpha_k = \sum_{i=0}^m \Lambda_{ki} u_i.
\end{equation}
Then Eq.~\eqref{kernel_eqn} shows that $\alpha_k=0$ whenever $c_k \neq 0$. If we denote by $\Upsilon_{ki} (\Psi)$ the submatrix of $\Lambda_{ki}$ where all rows with row label $k$ where $c_k=0$ are removed, then our derivation implies that $\uu \in \ker\Upsilon(\Psi)$.
Conversely, if $\uu \in \ker\Upsilon(\Psi)$ then it follows immediately from the equation above that $\uu \in \ker\Gram(\Psi)$.
We therefore find the equation
\begin{equation}
\ker\Gram(\Psi) = \ker\Upsilon(\Psi)
\end{equation}
as the main result of this appendix.
The $\Upsilon$ matrix defined here is exactly the same as in \citet[Sec.~III.B]{penz2021-Graph-DFT} that was used to characterize counterexamples to the Hohenberg--Kohn theorem on lattices. This shows that this previous investigation relied on the same concepts, although the Gram matrix or critical points were not defined there.

It is instructive to give a constructive description for critical values $\bb=\gmap(\Psi)\in\Bcrit$ where $\Psi\in\crit$. They are defined as coming from states $\Psi\in\M$ where the $\hat B_i\Psi$, $i=0,\ldots,m$, are linearly dependent. First, it is obvious that all $\Phi_k$ are critical points themselves, since $\hat B_i\Phi_k=\Lambda_{ki}\Phi_k$ always gives a vector parallel to $\Phi_k$. This means the $L$ vertices of $\Bdom$, $\gmap(\Phi_k)$, are all critical values. We then take any two such vertices and consider all convex combinations $\lambda\gmap(\Phi_k)+(1-\lambda)\gmap(\Phi_l)$, $\lambda\in[0,1]$, that are exactly the densities of the superpositions $\Psi=\sqrt{\lambda}\Phi_k+\sqrt{1-\lambda}\Phi_l$ since
\begin{equation}\begin{aligned}
    g_i(\Psi) = \langle \hat B_i \rangle_{\Psi} =&\; \lambda \langle \hat B_i \rangle_{\Phi_k} + (1-\lambda)\langle \hat B_i \rangle_{\Phi_l} \\&+ 2\sqrt{\lambda(1-\lambda)}\Re\langle \Phi_k,\hat B_i\Phi_l \rangle\\
    =&\; \lambda g_i(\Phi_k) + (1-\lambda)g_i(\Phi_l),
\end{aligned}\end{equation}
where the mixed term vanishes because $\Phi_k,\Phi_l$ are both eigenvectors of $\hat B_i$ and are further orthogonal. Now in the expansion with respect to the basis $\{\Phi_k\}$, these convex combinations have two non-zero coefficients while ${M\choose N}-2$ coefficients are zero. Surely, for $M> 2$, this will be not enough to make all $\hat B_i\Psi$ linearly independent. 
We can then proceed iteratively by mixing in a third basis state, etc. The question then arises as to how many non-zero coefficients we need to guarantee $\Psi\notin\crit$ or, asked differently, how many zero coefficients tell us that $\Psi\in\crit$. For shortened notation, call $\nu(\Psi)$ the number of non-zero coefficients in the given basis expansion. It was already noted that since the $\hat B_i$ are linearly independent, the columns of the matrix $\Lambda_{ki}$ are as well. Writing
\begin{equation}
    \hat B_i\Psi = \sum_{k=1}^L c_k\Lambda_{ki}\Phi_k,\quad i=0,\ldots,m
\end{equation}
then shows that the $\hat B_i\Psi$ are all linearly independent if the columns of the matrix $c_k\Lambda_{ki}$ are or, equivalently, this matrix has full rank $M$. Since every $c_k=0$ deletes one row while for $c_k\neq 0$ they are kept, we need $\nu(\Psi)\geq M$ for full rank. Conversely, this means that if $\nu(\Psi)\leq M-1$ then $\Psi\in\crit$ and $\gmap(\Psi)\in\Bcrit$. So convex combinations of up to $M-1$ vertices $\gmap(\Phi_k)$ are critical values.
A related argument was used in \citet[Sec.~III.B]{penz2021-Graph-DFT} to characterize counterexamples to the Hohenberg--Kohn theorem on lattices (see Appendix~\ref{app:crit} for this connection). There, an interesting theorem by \citet{Odlyzko} was further used to guarantee a full rank $M$ for the matrix $c_k\Lambda_{ki}$. This result can be used complementarily here to give a lower bound on $\nu(\Psi)$ that rules out $\Psi\in\crit$.
As an example, take a lattice with $M=4$ sites and $N=2$ particles, then the Hilbert-space dimension is $L={M\choose N}=6$. If $\nu(\Psi)\leq 3$ then $\Psi\in\crit$ for sure by the previous discussion, while by the result of \citet{Odlyzko} we need $\nu(\Psi)>4$ to infer $\Psi\notin\crit$. This means that in this example the value $\nu(\Psi)=4$ is left as undecided, some such $\Psi$ can be critical points, others are not.

\section{Proof of solvability for \texorpdfstring{$\bbeta$}{beta}}
\label{app:sol-beta}

In order to see that Eq.~\eqref{eq:constraints-beta-simple}, or more generally, Eq.~\eqref{eq:constraints-beta-simple-gen}, can always be solved for $\bbeta = (\beta_0,\ldots,\beta_{m})$ (in this appendix $\bbeta$ always has $M=m+1$ components), remember that the Gram matrix $\Gram=\Gram(\Psi)$ is positive definite and thus invertible if and only if all $\hat B_i\Psi$, $i=0,\ldots,m$, are linearly independent. This is by definition the case if $\Psi$ is a regular point, which in turn is guaranteed if $\bb=\gmap(\Psi)$ is a regular value. We then also have a \emph{unique} solution $\bbeta$ from Eq.~\eqref{eq:constraints-beta-simple-gen}, just as in the case of the Hohenberg--Kohn theorem in Appendix~\ref{app:gft}.

If, on the other hand, we have a critical point $\Psi\in\crit$ and consequently, $\Gram$ is \emph{not} positive definite, then $\Gram$ has a nontrivial kernel and we can choose a real vector $\vv\in\ker\Gram$, $\vv\neq 0$, that has
\begin{equation}
\sum_{j=0}^m \Gram_{ij} v_j=0
\end{equation}
for all $i=0,\ldots,m$. Then
\begin{equation}\label{eq:gamma-zero}
0= \sum_{i,j=0}^m v_i \Gram_{ij} v_j = \sum_{i,j=0}^m \langle v_i \hat{B}_i \Psi, v_j \hat{B}_j \Psi \rangle = \langle \chi, \chi \rangle,
\end{equation}
where we have defined
\begin{equation}
\chi = \sum_{i=0}^m v_i \hat{B}_i \Psi.
\end{equation}
Now take the vector $\bm{\gamma}$ from the right-hand side of Eq.~\eqref{eq:constraints-beta-simple}, $\Gram\cdot\bbeta=\bm{\gamma}$,  that has components $\gamma_i = -\Re \langle \hat{A} \Psi, \hat{B}_i \Psi \rangle$.
It follows from Eq.~\eqref{eq:gamma-zero} that $\chi=0$ and therefore
\begin{equation}
\sum_{i=0}^m v_i \gamma_i = -\Re \langle \hat{A} \Psi, \chi \rangle=0.
\label{gamma_ortho}
\end{equation}
Since $\vv$ was an arbitrary kernel vector we thus see that
\begin{equation}
\bm{\gamma} \in (\ker\Gram)^\perp = \im(\Gram^\top) =  \im\Gram,
\end{equation}
which implies that $\Gram\cdot\bbeta=\bm{\gamma}$ always has a solution (this argument corresponds to the Fredholm alternative). Yet, in the case of critical $\Psi$, this solution can be non-unique.

Let us further show that even for a critical $\Psi$ the evolution equation $-\partial_\tau \Psi(\tau) = \hat G(\tau)\Psi(\tau)$ is still well-defined. The non-unique solution $\bbeta$ at a critical $\Psi$ can be written as
\begin{equation}
\bbeta = \ww + \vv,
\end{equation}
where $\ww$ is the unique vector in $(\ker\Gram)^\perp$ such that $\Gram \cdot \ww=\bm{\gamma}$ and $\vv$ is an arbitrary vector in $\ker\Gram$.
Then writing out the generator from Eq.~\eqref{eq:generator-gen}, we have
\begin{equation}
    -\partial_\tau \Psi = \hat{A} \Psi + \sum_{i=0}^m (w_i+v_i) \hat{B_i} \Psi = \hat{A} \Psi + \sum_{i=0}^m w_i \hat{B_i} \Psi,
\end{equation}
since $\sum_{i=0}^M v_i \hat{B}_i \Psi=\chi=0$ according to Eq.~\eqref{eq:gamma-zero} above. Since $\ww$ was uniquely determined, it follows that the evolution equation is well-defined even when $\Psi$ is critical and $\bbeta$ cannot be uniquely determined.

\section{Solution to the evolution equation and the differentiability of \texorpdfstring{$\bbeta(\tau)$}{beta(tau)}}
\label{app:differentiability-beta}

We consider some general properties of our evolution equation \eqref{eq:evolution-nl} which, since $\H$ is finite dimensional, represents a first-order system
of autonomous ordinary differential equations (ODEs).
For this purpose, we rewrite it into the standard form of an ODE,
\begin{equation}
\partial_\tau \Psi = f(\Psi),
\label{ODE1}
\end{equation} 
where we define
\begin{equation}
 f(\Psi) := -\hat G(\Psi)\Psi = -\hat{A} \Psi - \sum_{i=0}^M \beta_{i} ( \Psi)  \hat{B}_i \Psi.
\end{equation}
Here, $\beta_i (\Psi)$ is defined to be the unique solution of the linear system $\Gram(\Psi)\cdot\bbeta=\bm{\gamma}$ if $\Psi \notin \crit$ as discussed in Appendix~\ref{app:sol-beta} (we later discuss how to relax this).
For a given initial state $\Psi_0 = \Psi (\tau=0)$ we will employ the Picard--Lindel\"of theorem~\cite{teschl-ode-book} to show that Eq.~\eqref{ODE1} has a unique solution.
For this purpose, one first converts the problem into an integral equation,
\begin{equation}
\Psi (\tau) = \Psi_0 + \int_{0}^\tau f(\Psi (s)) \d s,
\label{inteq}
\end{equation}
and then assumes that on a suitable domain $\Omega\subseteq\H$ the function $f(\Psi)$ is Lipschitz, i.e.,
\begin{equation}
\| f(\Psi_1) - f(\Psi_2) \| \leq \mathcal{L} \| \Psi_1 - \Psi_2 \|
\end{equation}
for some $\mathcal{L}>0$ and any $\Psi_1, \Psi_2 \in \Omega$. For $\Psi_0 \in \Omega$ the Picard--Lindel\"of theorem then establishes the existence of
a unique continuous solution to the integral equation on a finite (possibly short) time interval $[0,a]$, $a>0$. 
This solution is then seen to be differentiable with respect to $\tau$ and thus satisfies Eq.~\eqref{ODE1}.

Let us apply this now to our case, for which we take $\Omega$ to be a compact connected submanifold of $\M_{\bb} \setminus \crit$, i.e.,
a compact subset of the constraint manifold away from the critical set 
$\crit$.
For $f(\Psi)$ to be Lipschitz it is sufficient to prove that all $\beta_i(\Psi)$ have bounded partial derivatives with respect to the wavefunction components on the set $\Omega$.
The explicit solution for $\beta_i$ is, by Cramer's rule, given by
\begin{equation}
\beta_i (\Psi) = \frac{\det\Gram_i (\Psi)}{\det\Gram (\Psi)},
\label{beta_eq}
\end{equation}
where $\Gram_i (\Psi)$ is the matrix $\Gram(\Psi)$ in which column $i$ is replaced by the vector $\bm{\gamma}$ discussed in Appendix \ref{app:sol-beta}.
So $\beta_i(\Psi)$ is the ratio between two polynomials in the wavefunction coefficients. Since $\det\Gram(\Psi)$ never vanishes on $\Omega$, all partial derivatives with respect to the wavefunction coefficients exist and are finite. We conclude using the Picard--Lindel\"of theorem that there exists a unique solution $\Psi(\tau)$ to Eq.~\eqref{ODE1} that is a $C^1$-function within $\Omega$ on a sufficiently small interval $[0,a]$. Since $\Psi(\tau)$ is a $C^1$-function and $\hat B_i\Psi\neq 0$ for $\Psi\in\Omega$ (remember that $\Omega\cap\crit=\emptyset$), it follows by applying the chain rule to Eq.~\eqref{beta_eq} that also the $\beta_i(\tau)$ are $C^1$-functions on $[0,a]$. This justifies taking the $\tau$-derivatives of $\beta_i$ in Eqs.~\eqref{Hderv1} and \eqref{Hderv2}.

We thus have established that the ODE has a well-defined unique solution away from the critical points.
Let us now suppose that $\Psi_0 \in \crit$, then we know that at that point some of the $\hat B_i\Psi$ become linearly dependent. This corresponds to some kind of singularity in the constraint set $\M_\bb$, and we can therefore not expect $f( \Psi )$ to be differentiable at $\Psi_0$. However, we will show that $f(\Psi)$ is still continuous at a critical point, which allows us to apply Peano's theorem~\cite{teschl-ode-book}, which still guarantees a $C^1$-solution
of the ODE through the critical point, albeit not a unique one.  
To do so, we show that 
\begin{equation}
\lim_{\epsilon \rightarrow 0} f(\Psi_0  + \epsilon \Phi) = f(\Psi_0)
\end{equation} 
in all directions $\Phi\in\H$.
For later reference, we denote by $\Psi_\epsilon := \Psi_0 + \epsilon \Phi$ a line segment in $\H$ and further define $\Gram_\epsilon := \Gram (\Psi_\epsilon)$ and $\bm{\gamma}_\epsilon := \bm{\gamma} (\Psi_\epsilon)$.
We can write out explicitly
\begin{align}
\Gram_\epsilon &=  \Gram (  \Psi_0 ) + \epsilon \Gram^{(1)} + \epsilon^2 \Gram (  \Phi ), \\
\bm{\gamma}_\epsilon &= \bm{\gamma} (\Psi_0) + \epsilon \bm{\gamma}^{(1)} + \epsilon^2 \bm{\gamma} (\Phi),
\end{align}
where we additionally defined
\begin{align}
\Gram_{ij}^{(1)} &:= 2\Re\langle \hat{B}_i\Psi_0, \hat{B}_j \Phi \rangle ,
\label{G1}\\
\gamma_i^{(1)} &:= - \Re( \langle \hat {A} \Psi_0, \hat{B}_i \Phi \rangle + \langle \hat {A} \Phi, \hat{B}_i \Psi_0 \rangle ).
\end{align}
From $\Psi_0\in\crit$ we have $\det\Gram ( \Psi_0 )=0$, while due to positive definiteness of the Gram matrix we have $\det\Gram_\epsilon \geq 0$. Since $\det \Gram_\epsilon$ is clearly analytic in $\epsilon$, it thus must have a zero of at least order 2 at $\epsilon=0$. Taken together, this means
\begin{equation}
     \det\Gram_\epsilon = c \, \epsilon^{2p} + O (\epsilon^{2p+1})
\end{equation}
with $c > 0$ and $p \geq 1$, or, if $\Gram_\epsilon$ has a zero eigenvalue independent of $\epsilon$, then
$\det\Gram_\epsilon =0$ for all values of $\epsilon$.
We now consider for $\epsilon >0$ the equation
\begin{equation}
\Gram_\epsilon \cdot \bm{\beta}_\epsilon = \bm{\gamma}_\epsilon
\label{eps_eq}
\end{equation}
along the line segment $\Psi_\epsilon$.
Denote the eigenvalues and eigenvectors of $\Gram_\epsilon$ as $\lambda_j (\epsilon)\in\R_{\geq 0}$ and $\uu_j (\epsilon)\in\R^{m+1}$.
Since $\Gram_\epsilon$ is analytic in $\epsilon$ it follows from Rellich's theorem~\cite{rellich1937,rellich-book} that there exist
eigenvalues $\lambda_j (\epsilon)$ and eigenvectors $\uu_j (\epsilon)$ which are analytic functions of $\epsilon$ around $\epsilon=0$. In his book, \citet[Ch.~1, {\S}1]{rellich-book} stresses that the unperturbed set of eigenvectors (i.e., at $\epsilon=0$) may not be prescribed in advance, since generally only very specific eigenvectors at $\epsilon=0$ can be connected in a smooth way to the ones at finite $\epsilon$ (see \citet[Ch.~1, {\S}1]{rellich-book} for an illustrative example). Nevertheless, Rellich's theorem importantly ensures that such a choice of vectors that are analytic in $\epsilon$ is always possible. Their real and imaginary parts are then analytic as well and can be normalized to yield $\mathbf{u}_j (\epsilon)$ as the (real and analytic) eigenvectors of the real and symmetric matrix $\Gram (\Psi_0)$.
We will further divide the eigenvalues of $\Gram_\epsilon$ into three sets
(some of which may be empty). The first set contains the eigenvalues that
are identically zero and independent of $\epsilon$, the second set contains the eigenvalues that depend on $\epsilon$ but become zero in the limit $\epsilon \rightarrow 0$, and the third set contains the eigenvalues that are non-zero even in the limit $\epsilon \rightarrow 0$. The collection of labels $i$
of $\lambda_i (\epsilon)$ of those three sets will be denoted by $A$, $B$, and $C$, respectively.
The solution to Eq.~\eqref{eps_eq} can then be expanded as 
\begin{equation}
 \bbeta_\epsilon = \vv (\epsilon) + \ww (\epsilon) + \mathbf{y},
 \label{beta_sol}
\end{equation}
where $\mathbf{y}$ is an arbitrary linear combination of eigenvectors with labels in $A$ and is therefore in the kernel of $\Gram_\epsilon$ and independent of $\epsilon$, while
\begin{align}
\vv (\epsilon) &=   \sum_{j\in B} \frac{ \uu_j (\epsilon) \cdot \bm{\gamma}_\epsilon}{\lambda_j (\epsilon)} \uu_j (\epsilon),  
\label{vv_eq}\\
\ww (\epsilon) &=   \sum_{j\in C} \frac{ \uu_j (\epsilon) \cdot \bm{\gamma}_\epsilon }{\lambda_j (\epsilon)} \uu_j (\epsilon). 
\end{align}
Since for $j \in C$ we have $\lambda_j (0) \neq 0$, the vector $\ww (0)$ is the well-defined limit of $\ww(\epsilon)$ for $\epsilon \rightarrow 0$.
Let us therefore consider the eigenvalues with labels $j\in B$.
For these eigenvalues and the corresponding eigenvectors we have the expansions
\begin{align}
\lambda_j (\epsilon) & = \epsilon^\ell \, \lambda_j^{(\ell)} + O (\epsilon^{\ell+1}), \\
\uu_j (\epsilon) & = \uu_j (0) + \epsilon^k \, \uu_j^{(k)} + O (\epsilon^{k+1}).
\label{u_order}
\end{align}
Here, for a given $j$, the integers $\ell$ and $k$ denote the orders of lowest
non-vanishing coefficients and coefficient vectors beyond the zeroth-order terms in the expansions. 
We continue by deriving some conditions on the values of $\ell$ and $k$.
Since $\Gram_\epsilon$ is positive semi-definite we have $\lambda_j (\epsilon) \geq 0$, and since the eigenvalues are also analytic at $\epsilon=0$ it follows that
$\ell \geq 2$ is even since $\lambda_j (\epsilon)$ cannot cross zero.
By expanding the eigenvalue equation $\Gram_\epsilon \cdot \uu_j (\epsilon) 
= \lambda_j (\epsilon) \uu_j (\epsilon)$ in powers of $\epsilon$ we 
find
\begin{align}
&[ \epsilon^2 \Gram (\Phi)\cdot \uu_j (0) + \epsilon^k \Gram (\Psi_0) \cdot \uu_j^{(k)} \nonumber \\
&+ \epsilon^{k+1}
(\Gram^{(1)} \cdot \uu_j^{(k)}+ \Gram (\Psi_0) \cdot \uu_j^{(k+1)})+ O(\epsilon^{k+2})] \nonumber \\
 &= \epsilon^\ell \lambda_j^{(\ell)} \uu_j (0)+ O (\epsilon^{\ell+1}),
 \label{eigen_exp}
 \end{align}
where we used that for $j\in B$ we have $\Gram (\Psi_0) \cdot \uu_j (0) =0$ as well as $\Gram^{(1)} \cdot \uu_j (0)=0$, which follows from an explicit calculation using Eq.~\eqref{G1}.
We will next show that the only option is $\ell=2$.
To do so, we note that $\Gram (\Phi)\cdot \uu_j (0)$ is non-zero because 
otherwise $\Gram_\epsilon \cdot \uu_j (0)=0$ for all $\epsilon$
and $\uu_j (0)$ would be an $\epsilon$-independent vector in the kernel of $\Gram_\epsilon$, which corresponds to the case in which $j \in A$, which
we treated already separately. Therefore the left-hand side of Eq.~\eqref{eigen_exp} at least contains a non-zero term of order 2. If $\ell > 2$
then Eq.~\eqref{eigen_exp} can only be satisfied if the lowest two orders on the left-hand side of the equation vanish,
\begin{align}
  &\epsilon^2 \Gram (\Phi)\cdot \uu_j (0) + \epsilon \, \Gram (\Psi_0) \cdot \uu_j^{(1)}   \nonumber \\
   &+\epsilon^2 ( \Gram^{(1)} \cdot \uu_j^{(1)}+ \Gram (\Psi_0) \cdot \uu_j^{(2)}) = 0.
\end{align}
This in turn requires that the first-order term vanishes, $\Gram (\Psi_0) \cdot \uu_j^{(1)}=0$. Since we also have $\Gram^{(1)} \cdot \uu_j^{(1)}=0$ like noted after Eq.~\eqref{eigen_exp}, we arrive at the condition
\begin{equation}
    \Gram (\Phi)\cdot \uu_j (0) = - \Gram (\Psi_0) \cdot \uu_j^{(2)}.
\end{equation}
But then by taking the dot product from the left with $\uu_j (0)$ and using again that $\uu_j (0)\in \ker \Gram (\Psi_0)$ for $j\in B$ we find
\begin{equation}
    \uu_j (0) \cdot  \Gram (\Phi)\cdot \uu_j (0) = 0.
\end{equation}
Since $\Gram (\Phi)$ is positive semi-definite, this implies $\Gram (\Phi) \cdot \uu_j (0)=0$, which is contradictory to our assumption that $j\in B$ as noted before. This shows that $\ell > 2$ is not possible, which implies that $\ell=2$. It follows from Eq.~\eqref{eigen_exp} that $k=1$ is possible but only if $\Gram (\Psi_0) \cdot \uu_j^{(1)}=0$, so $\uu_j^{(1)} \in \ker \Gram (\Psi_0)$, otherwise we have $k \geq 2$.
We can now collect our results and study the $\epsilon \rightarrow 0$ limit of the coefficients in Eq.~\eqref{vv_eq}. We start by expanding
\begin{equation}
    \uu_j (\epsilon) \cdot \bm{\gamma}_\epsilon = a_j \epsilon  + b_j \epsilon^2 +  c_j \epsilon^k+ O(\epsilon^{k+1}),
\end{equation}
where we defined
\begin{align}
a_j &=  \uu_j (0) \cdot \bm{\gamma}^{(1)}, \\
b_j &=  \uu_j (0) \cdot \bm{\gamma} (\Phi), \\
c_j &=  \uu_j^{(k)} \cdot \bm{\gamma} (\Psi_0),
\end{align}
and used $\uu_j (0) \cdot \bm{\gamma} (\Psi_0)=0$ (see Eq.~\eqref{gamma_ortho}).
For the case $(\ell,k)=(2,1)$ we have 
 \begin{equation}
     \frac{ \uu_j (\epsilon) \cdot \bm{\gamma}_\epsilon}{\lambda_j (\epsilon)} \uu_j (\epsilon)
     = \left(\frac{a_j}{\lambda_j^{(2)} \epsilon }+ O (1)\right) \uu_j (0) + \frac{a_j}{\lambda_j^{(2)}} \uu_j^{(1)}+ O (\epsilon),
     \label{lk_21}
 \end{equation}
where we additionally used that $c_j=0$ from Eq.~\eqref{gamma_ortho} since $\uu_j^{(1)} \in \ker \Gram (\Psi_0)$.
Finally, for the remaining case $(\ell,k)=(2,k)$ with $k \geq 2$ we have 
 \begin{equation}
     \frac{ \uu_j (\epsilon) \cdot \bm{\gamma}_\epsilon}{\lambda_j (\epsilon)} \uu_j (\epsilon)
     = \left(\frac{a_j}{ \lambda_j^{(2)}\epsilon }+ O (1)\right) \uu_j (0) + O (\epsilon).
     \label{lk_2k}
 \end{equation}
 In Eqs.~\eqref{lk_21} and \eqref{lk_2k} the singular and constant part of the expansion on the right-hand sides
 are all in the kernel of $\Gram (\Psi_0)$. 
 Since for any $\vv \in \ker \Gram(\Psi_0)$ we have that $\sum_{i=0}^m v_i \hat{B}_i \Psi_0=0$ (see Appendix \ref{app:sol-beta})
 it therefore follows that
 \begin{equation}
     \lim_{\epsilon \rightarrow 0} \sum_{i=0}^m v_i (\epsilon) \hat{B}_i \Psi_\epsilon = 0.
 \end{equation}
Since in Eq.~\eqref{beta_sol} also $\mathbf{y} \in \ker \Gram (\Psi_0)$ and, as explained in Appendix~\ref{app:sol-beta}, the values $w_i (0)$ are uniquely determined by $\Psi_0$ the limit
\begin{equation}
\lim_{\epsilon \rightarrow 0} \sum_{i=0}^m \beta_i  (\Psi_\epsilon ) \hat{B}_i \Psi_\epsilon =  \sum_{i=0}^m w_i (0)  \hat{B}_i \Psi_0 
\end{equation}
is well-defined. It thus follows that $\lim_{\epsilon \rightarrow 0} f(\Psi_\epsilon) = f(\Psi_0)$.

This result now allows us to apply Peano's theorem \cite{teschl-ode-book} to establish that there is a solution to the ODE passing through the singular point. Due to a theorem by \citet{Kneser1923} it follows that in case of non-uniqueness there will be a whole continuum of such solutions solving the initial-value problem passing through this point.

From the analysis in this appendix the following picture arises. Either we have a full solution to our initial-value problem which exists for all times without ever reaching a critical point (including the case in which a critical point is approached for $\tau \rightarrow \infty$), or we reach a critical point at a finite time and continuation through the critical point exists but is not unique. The time evolution can then be continued to all times or we may pass in a non-unique way through more critical points. In any case, we find that a global solution always exists but may not be unique.

\newpage

%

\end{document}